\documentclass[trackchanges, twocolumn,resetfootnote]{aastex701}

\usepackage{rotating}

\begin{document}

\title{RIOJA. Environmental Effects on Stellar Populations and Ionized Gas in a Protocluster at $z=7.88$}

\correspondingauthor{Wataru Osone}
\author[orcid=0009-0008-8424-100X, sname='Wataru']{Wataru Osone}
\affiliation{Division of Physics, Faculty of Pure and Applied Sciences, University of Tsukuba, Tsukuba, Ibaraki 305-8571, Japan}
\email[show]{s2420153@u.tsukuba.ac.jp}

\author[orcid=0000-0002-0898-4038, sname='Takuya']{Takuya Hashimoto} 
% \altaffiliation{University of Tsukuba}
\affiliation{Division of Physics, Faculty of Pure and Applied Sciences, University of Tsukuba, Tsukuba, Ibaraki 305-8571, Japan}
\affiliation{Tsukuba Institute for Advanced Research (TIAR), University
of Tsukuba, 1-1-1 Tennodai, Tsukuba, Ibaraki, 305-8577, Japan}
\affiliation{Tomonaga Center for the History of the Universe (TCHoU), Faculty of Pure and Applied Sciences, University of Tsukuba, Tsukuba, Ibaraki 305-8571, Japan}
\email[show]{hashimoto.takuya.ga@u.tsukuba.ac.jp}

\author[orcid=0000-0002-7093-1877, sname='Javier']{Javier Álvarez-Márquez} 
\affiliation{Centro de Astrobiología (CAB), CSIC-INTA, Ctra. de Ajalvir km 4, Torrejón de Ardoz, E-28850, Madrid, Spain}
\email{jalvarez@cab.inta-csic.es}

\author[orcid=0000-0002-9090-4227, sname='Luis']{Luis Colina} 
\affiliation{Centro de Astrobiología (CAB), CSIC-INTA, Ctra. de Ajalvir km 4, Torrejón de Ardoz, E-28850, Madrid, Spain}
\email{colina@cab.inta-csic.es}

\author[orcid=0000-0002-7779-8677, sname='Akio']{Akio K. Inoue} 
\affiliation{Waseda Research Institute for Science and Engineering, Faculty of Science and Engineering, Waseda University, 3-4-1 Okubo, Shinjuku, Tokyo 169-8555, Japan}
\affiliation{Department of Pure and Applied Physics, School of Advanced Science and Engineering, Faculty of Science and Engineering, Waseda University, 3-4-1 Okubo,
Shinjuku, Tokyo 169-8555, Japan}
\email{akinoue@aoni.waseda.jp}

\author[orcid=0000-0003-4985-0201, sname='Ken']{Ken Mawatari} 
\affiliation{Waseda Research Institute for Science and Engineering, Faculty of Science and Engineering, Waseda University, 3-4-1 Okubo, Shinjuku, Tokyo 169-8555, Japan}
\affiliation{Department of Pure and Applied Physics, School of Advanced Science and Engineering, Faculty of Science and Engineering, Waseda University, 3-4-1 Okubo,
Shinjuku, Tokyo 169-8555, Japan}
\email{mawatari@aoni.waseda.jp}

\author[orcid= 0000-0001-6958-7856, sname='Yuma']{Yuma Sugahara} 
\affiliation{Waseda Research Institute for Science and Engineering, Faculty of Science and Engineering, Waseda University, 3-4-1 Okubo, Shinjuku, Tokyo 169-8555, Japan}
\affiliation{Department of Pure and Applied Physics, School of Advanced Science and Engineering, Faculty of Science and Engineering, Waseda University, 3-4-1 Okubo,
Shinjuku, Tokyo 169-8555, Japan}
\email{sugayu@aoni.waseda.jp}

\author[orcid=0009-0005-5448-5239, sname='Carmen']{Carmen Blanco-Prieto} 
\affiliation{Centro de Astrobiología (CAB), CSIC-INTA, Ctra. de Ajalvir km 4, Torrejón de Ardoz, E-28850, Madrid, Spain}
\email{cblanco@cab.inta-csic.es}

\author[orcid=, sname='Takeshi']{Takeshi Hashigaya} 
\affiliation{Department of Astronomy, Kyoto University Sakyo-ku, Kyoto 606-8502, Japan}
\email{hashigaya@kusastro.kyoto-u.ac.jp}

\author[orcid=0000-0002-8512-1404, sname='Takahiro']{Takahiro Morishita} 
\affiliation{Astronomical Institute, Graduate School of Science, Tohoku University, Sendai, Miyagi 980-8578, Japan}
\email{morishita@astr.tohoku.ac.jp}

\author[orcid=0000-0001-9935-6047, sname='Massimo']{Massimo Stiavelli} 
\affiliation{Space Telescope Science Institute, 3700 San Martin Drive, Baltimore, MD 21218, USA}
\affiliation{The William H. Miller III, Dept. of Physics \& Astronomy, Johns Hopkins University, Baltimore, MD 21218, USA}
\email{mstiavel@stsci.edu}

\author[orcid=0000-0001-7997-1640, sname='Santiago']{Santiago Arribas} 
\affiliation{Centro de Astrobiología (CAB), CSIC-INTA, Ctra. de Ajalvir km 4, Torrejón de Ardoz, E-28850, Madrid, Spain}
\email{arribas@cab.inta-csic.es}

\author[orcid=0000-0002-5268-2221, sname='Tom']{Tom J. L. C. Bakx} 
\affiliation{Department of Physics, Graduate School of Science, Nagoya University, Nagoya, Aichi 464-8602, Japan}
\affiliation{Department of Space, Earth, \& Environment, Chalmers University of Technology, Chalmersplatsen 4 412 96 Gothenburg, Sweden}
\affiliation{National Astronomical Observatory of Japan, 2-21-1, Osawa, Mitaka, Tokyo, Japan}
\email{tjlcbakx@gmail.com}

\author[orcid=0000-0002-8680-248X, sname='Daniel']{Daniel Ceverino} 
\affiliation{Departamento de Fisica Teorica, Modulo 8, Facultad de Ciencias, Universidad Autonoma de Madrid, 28049 Madrid, Spain}
\affiliation{CIAFF, Facultad de Ciencias, Universidad Autonoma de Madrid, 28049 Madrid, Spain}
\email{daniel.ceverino@uam.es}

\author[orcid=0000-0003-2119-277X, sname='Alejandro']{Alejandro Crespo Gómez} 
\affiliation{Space Telescope Science Institute, 3700 San Martin Drive, Baltimore, MD 21218, USA}
\email{acrespo@stsci.edu}

\author[orcid=0000-0001-7440-8832, sname='Yoshinobu']{Yoshinobu Fudamoto} 
\affiliation{Center for Frontier Science, Chiba University, 1-33 Yayoi-cho, Inage-ku, Chiba 263-8522, Japan}
\email{yoshinobu.fudamoto@gmail.com}

\author[orcid=0000-0001-8083-5814, sname='Masato']{Masato Hagimoto} 
\affiliation{Department of Physics, Graduate School of Science, Nagoya University, Nagoya, Aichi 464-8602, Japan}
\email{hagimoto@a.phys.nagoya-u.ac.jp}

\author[orcid=, sname='Asahi']{Asahi Hamada} 
\affiliation{Division of Physics, Faculty of Pure and Applied Sciences, University of Tsukuba, Tsukuba, Ibaraki 305-8571, Japan}
\email{s2520200@u.tsukuba.ac.jp}

\author[orcid=0000-0001-8442-1846, sname='Rui']{Rui Marques-Chaves} 
\affiliation{Geneva Observatory, Department of Astronomy, University of Geneva, Chemin Pegasi 51, CH-1290 Versoix, Switzerland}
\email{rui.marquescoelhochaves@unige.ch}

\author[orcid=0000-0002-0984-7713, sname='Yurina']{Yurina Nakazato} 
\affiliation{Center for Computational Astrophysics, Flatiron Institute, 162 5th Avenue, New York, NY, 10010, USA}
\email{ynakazato@flatironinstitute.org}

\author[orcid=0000-0003-1937-0573, sname=Umehata]{Hideki Umehata}
\affiliation{Institute for Advanced Research, Nagoya University, Furocho, Chikusa, Nagoya 464-8602, Japan}
\affiliation{Department of Physics, Graduate School of Science, Nagoya University, Nagoya, Aichi 464-8602, Japan}
\email{umehata@a.phys.nagoya-u.ac.jp}  

\author[orcid=0000-0002-1319-3433, sname='Hidenobu']{Hidenobu Yajima} 
\affiliation{Center for Computational Sciences, University of Tsukuba, Ten-nodai, 1-1-1 Tsukuba, Ibaraki 305-8577, Japan}
\email{yajima@ccs.tsukuba.ac.jp}

\author[orcid=0000-0003-3278-2484, sname='Hiroshi']{Hiroshi Matsuo} 
\affiliation{National Astronomical Observatory of Japan, 2-21-1 Osawa, Mitaka, Tokyo 181-8588, Japan}
\affiliation{The Graduate University for Advanced Studies (SOKENDAI), 2-21-1 Osawa, Mitaka, Tokyo 181-8588, Japan}
\email{h.matsuo@nao.ac.jp}

\author[orcid=0000-0001-7925-238X, sname='Naoki']{Naoki Yoshida} 
\affiliation{Department of Physics, The University of Tokyo, 7-3-1 Hongo, Bunkyo, Tokyo 113-0033, Japan}
\affiliation{Kavli Institute for the Physics and Mathematics of the Universe (WPI), UT Institute for Advanced Study, The University of Tokyo, Kashiwa, Chiba 277-8583,
Japan}
\affiliation{Research Center for the Early Universe, School of Science, The University of Tokyo, 7-3-1 Hongo, Bunkyo, Tokyo 113-0033, Japan}
\email{naoki.yoshida@ipmu.jp}

\author[orcid=0000-0002-6510-5028, sname='Yi']{Yi W. Ren} 
\affiliation{Department of Pure and Applied Physics, School of Advanced Science and Engineering, Faculty of Science and Engineering, Waseda University, 3-4-1 Okubo, Shinjuku, Tokyo 169-8555, Japan}
\email{renyi@toki.waseda.jp}

\author[orcid=0000-0003-4807-8117, sname='Yoichi']{Yoichi Tamura} 
\affiliation{Department of Physics, Graduate School of Science, Nagoya University, Nagoya, Aichi 464-8602, Japan}
\email{ytamura@nagoya-u.jp}

\begin{abstract}
Protoclusters in the epoch of reionization provide key laboratories for investigating how environment shapes early galaxy formation and evolution, and may also have contributed to cosmic reionization.
We analyze 23 member galaxies of A2744-z7p9OD, a protocluster at $z=7.88$, using JWST/NIRCam and NIRSpec to investigate their stellar population properties, rest-frame UV sizes, and ionized-gas properties. We also quantify the internal structure of A2744-z7p9OD using the projected distance to the most massive galaxy ($D_{\rm YD4}$), and to the nearest neighbor ($D_{\rm nei}$), as global and local environmental indicators, respectively.
Stellar mass, SFR on a 100 Myr timescale, dust attenuation, and galaxy size show significant correlations ($p<0.05$) with $D_{\rm YD4}$, but not with $D_{\rm nei}$, suggesting that these properties are primarily linked to the global protocluster structure. The member galaxies also show a large galaxy-to-galaxy variation in R23 ($=\log{(([\mathrm{O}\text{\textsc{iii}}]\lambda\lambda4960,5008\rm{\AA}+[\mathrm{O}\text{\textsc{ii}}]\lambda\lambda3727,3730\rm{\AA})/\rm{H}\beta)}$), implying inhomogeneous chemical enrichment in the protocluster environment.
O32 ($=\log{([\mathrm{O}\text{\textsc{iii}}]\lambda5008\rm{\AA}/[\mathrm{O}\text{\textsc{ii}}]\lambda\lambda3727,3730\rm{\AA})}$) correlates with $D_{\rm YD4}$, indicating that the core region is characterized by low-ionization gas. Together with the non-detection of Ly$\alpha$ emission, the possible neutral-gas reservoir traced by ALMA [C{\sc ii}]~$158\mu$m emission, and evidence for high-column-density neutral hydrogen in the core, this suggests a neutral-gas-rich protocluster core where the current escape of ionizing photons may be suppressed, even in a overdense environment during the EoR.

\end{abstract}


\section{Introduction} 
Overdense environments in the local Universe host a large fraction of evolved galaxies and galaxy properties strongly depend on environment (e.g., \citealt{Dressler1980mar}). Such dependence of galaxy properties on their surrounding environment is commonly referred to as environmental effects. 
To understand how environment shapes galaxy formation and evolution, it is necessary to study protoclusters, i.e., high-redshift galaxy overdensities that are expected to evolve into present-day galaxy clusters.

In protocluster environments, galaxy--galaxy interactions, mergers, gas accretion, star formation, metal enrichment, and dust build-up can proceed differently from those in the field (e.g., \citealt{Overzier2016nov}; \citealt{Alberts2022oct}). 
Observational studies of protoclusters at $z\sim2-4$ have shown that environmental effects can influence galaxy properties in complex ways. The stellar mass function in a protocluster at $z\sim3.3$ shows an enhanced fraction of massive galaxies in the overdense region compared with the field, suggesting accelerated stellar mass assembly in dense environments \citep{Forrest2024aug}. Gas-phase metallicity shows a more complex behavior: protocluster environments may promote chemical enrichment through enhanced galaxy growth (e.g., \citealt{Shimakawa2015mar}), but cold gas inflows can also dilute the metal content of galaxies in overdense regions (e.g., \citealt{Calabro2022aug}). Atacama Large Millimeter/submillimeter Array (ALMA; \citealt{Wootten2009aug}) observations have further revealed concentrations of dusty star-forming galaxies and molecular-gas-rich galaxies in protoclusters, indicating enhanced gas supply and dust-obscured star formation (e.g., \citealt{Umehata2015dec, Tadaki2019apr}). Because such environmental dependence is linked to galaxy mass and size, surrounding gas reservoir, the galaxy spatial distributions, and the evolutionary stage of the protocluster, it is important to constrain both the stellar population and interstellar medium (ISM) properties of member galaxies and examine how they are connected to the protocluster environment.

Protoclusters in the epoch of reionization (EoR) provide a particularly important testbed. At this epoch, early sites of cluster formation were beginning to emerge as galaxy overdensities, and such environments may have already shaped early galaxy evolution (e.g., \citealt{Chiang2017aug}). Protoclusters in the EoR may also have contributed to cosmic reionization because numerous galaxies in dense environments can collectively produce ionizing photons and form large ionized bubbles (e.g., \citealt{Furlanetto2004sep, McQuinn2007may, Castellano2016feb, Endsley2022apr}).

The advent of James Webb Space Telescope (JWST; \citealt{Gardner2023jun}) has greatly advanced studies of protoclusters in the EoR, enabling both the discovery of member galaxies and detailed investigations of their stellar population and ISM properties through deep NIRCam imaging, NIRCam wide-field slitless spectroscopy (WFSS), and NIRSpec spectroscopy (e.g., \citealt{Morishita2023apr, Arribas2024aug, Helton2024oct, Witstok2025jan, Fudamoto2025mar, Witten2025feb, Li2026apr}).

Despite this progress, testing environmental effects in protoclusters remains challenging. This is particularly true in the EoR, where member galaxies are faint and spectroscopically confirmed samples are still limited. To understand protoclusters as physical environments that affect galaxy evolution, it is necessary to combine measurements of the physical properties of individual member galaxies with environmental indicators that quantify their locations within the protocluster.

Galaxies near the protocluster center reside in denser and more evolved environments, and may show signatures of earlier stellar mass assembly, changes in star formation activity, enhanced metal enrichment, or stronger dust attenuation. On the other hand, the distance to the nearest neighboring galaxy can trace local environments related to close galaxy–galaxy interactions and mergers. It is therefore essential to separately define global environmental indicators, such as the distance from the protocluster center, and local indicators, such as the nearest-neighbor distance, and to compare them with the physical properties of member galaxies.

Such environmental indicators have already been used in studies of protoclusters at $z \sim 2-7$. For example, \cite{Perez-Martinez2023jan} investigated the relation between star formation activity and environment in the Spiderweb protocluster at $z=2.16$, and \cite{Toshikawa2025mar} examined the relation between the spatial distribution of galaxies and their physical properties from the outskirts to the core in a protocluster at $z \sim 3.7$. \cite{Toshikawa2025mar} found that galaxy properties are more strongly correlated with the distance to the nearest neighboring galaxy than with the distance from the protocluster center. \citet{Champagne2025mar} applied a similar analysis to a quasar-anchored protocluster at $z=6.6$ using JWST data. In contrast, similar studies remain limited for protoclusters at $z>7$, mainly because only a small number of member galaxies have been spectroscopically confirmed in individual systems and because datasets that can simultaneously constrain stellar populations and ISM properties have been limited.

\begin{figure*}[t]
    \centering
    \includegraphics[width=0.9\textwidth,clip]{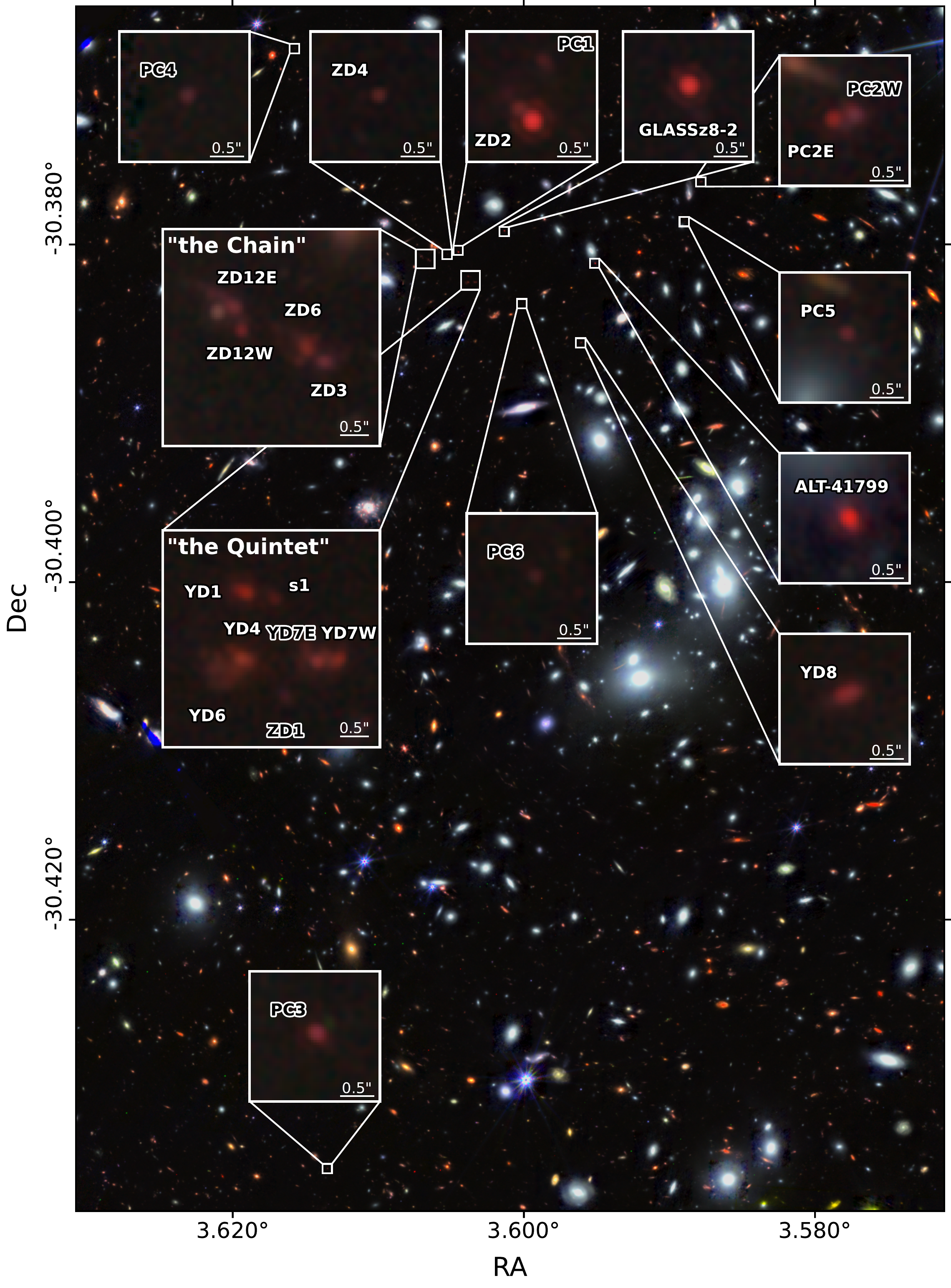}
    \caption{RGB composite image of A2744-z7p9OD constructed from JWST/NIRCam F115W (blue), F277W (green), and F444W (red) imaging. The inset panels show the member galaxies analyzed in this study. White labels indicate spectroscopically confirmed member galaxies, while black labels with white outlines indicate member candidates without spectroscopic redshift confirmation.}
    \label{fig:A2744z7p9OD}
\end{figure*}

In this context, A2744-z7p9OD is an ideal target. It is a galaxy overdensity at $z=7.88$ identified as a protocluster expected to evolve into a present-day massive galaxy cluster \citep{Morishita2023apr}. A2744-z7p9OD is one of the most extreme protoclusters in the early Universe, in terms of its high redshift, compact core, and the evolved properties of its member galaxies \citep{Atek2014may, Zheng2014nov, Laporte2014feb, Ishigaki2016may, Roberts-Borsani2022oct, Morishita2023apr, Cameron2024oct, Chen2024mar, Venturi2024nov, Morishita2025may, Witten2025feb, Witten2025jul}. The galaxy number overdensity, $\delta$, is estimated to be $\sim60$--$70$ \citep{Witten2025jul}.
Figure \ref{fig:A2744z7p9OD} shows the RGB image of the member galaxies in A2744-z7p9OD. 
This system hosts two compact galaxy associations in its core: ``the Quintet'', a compact association of at least five galaxies (e.g., \citealt{Hashimoto2023sep,Venturi2024nov,Fudamoto2025oct,Umehata2025nov}), and ``the Chain'', an elongated association of four member galaxies (e.g., \citealt{Morishita2025may,Umehata2025nov}). A total of 16 member galaxies have been spectroscopically confirmed with NIRSpec and NIRCam/WFSS, while seven additional galaxies have been reported as photometric member galaxy candidates \citep{Morishita2023apr,Witten2025jul}.
In addition, this protocluster is magnified by Abell 2744 by a factor of $\mu \sim 2$, making it a favorable target for detailed studies of the stellar populations and ISM of protocluster member galaxies during the EoR.

A2744-z7p9OD has been the target of many observational studies with JWST and ALMA. Despite its high redshift of $z \sim 7.88$, ALMA observations have detected dust continuum emission from three galaxies in ``the Quintet'' region \citep{Hashimoto2023sep, Fudamoto2025oct, Umehata2025nov}, as well as [C{\sc ii}] $158\,\mu$m in both ``the Quintet'' and ``the Chain'' \citep{Fudamoto2025oct, Umehata2025nov}. In particular, [C{\sc ii}] $158\,\mu$m peaks associated with YD1, YD4, and YD7W have been identified in ``the Quintet'' region, suggesting the presence of a neutral-gas reservoir around the protocluster core \citep{Fudamoto2025oct}. 
Deep JWST spectroscopy has also revealed damped Ly$\alpha$ absorption or strong Ly$\alpha$-break damping in some member galaxies of A2744-z7p9OD, providing further evidence for high-column-density neutral hydrogen \citep{Chen2024mar,Vasan2026jun}. JWST/NIRCam studies have also suggested that the core galaxies are massive, dusty, and show declining star formation activity, whereas galaxies in the outer regions are younger and have experienced more recent starbursts \citep{Witten2025jul}. These results indicate that A2744-z7p9OD may already host a highly evolved protocluster core at $z>7$. 

In this study, we combine JWST/NIRCam imaging with JWST/NIRSpec spectroscopy to perform a homogeneous analysis of the stellar populations and ionized-gas properties of the member galaxies in A2744-z7p9OD. We further connect these physical properties with structural indicators within the protocluster to investigate how the protocluster environment affects galaxy evolution at $z \sim 8$. 
We analyze 23 galaxies associated with A2744-z7p9OD, including 16 spectroscopically confirmed member galaxies and seven photometric member candidates, all within the central 6 cMpc of the protocluster (Table \ref{tab:spec_obs}).

This paper is organized as follows. Section~\ref{sec:Data} describes the data used in this study and their reduction. Section~\ref{sec:SED_Fitting} presents the stellar population properties of the member galaxies derived from Spectral Energy Distribution (SED) fitting. In Section~\ref{sec:Galaxy_sizes}, we investigate the sizes and star formation rate surface densities of the member galaxies.
Section~\ref{sec:Line_Analyses} investigates the ionized-gas properties of the member galaxies based on emission-line analyses. Section~\ref{sec:Protocluster_Structure} quantitatively characterizes the internal structure of A2744-z7p9OD. Section~\ref{sec:Discussion} discusses the impact of the dense environment during the EoR on galaxy evolution and the contribution of A2744-z7p9OD to cosmic reionization.
Throughout this paper, we adopt a $\Lambda$CDM cosmology, with $H_0 = 70\ \mathrm{km\ s^{-1}\ Mpc^{-1}}$, $\Omega_{M} = 0.3$, and $\Omega_{\Lambda} = 0.7$. Hereafter, unless otherwise noted, we refer to [O{\sc ii}] $\lambda\lambda$3727, 3730 $\mathrm{\AA}$ and [O{\sc iii}] $\lambda\lambda$4960, 5008 $\mathrm{\AA}$ simply as [O{\sc ii}]3727, 3730 and [O{\sc iii}]4960, 5008, respectively.

\begin{deluxetable*}{ccccccccc}
% \digitalasset %提出時は復活？
% \tablewidth{0pt}
\setlength{\tabcolsep}{3pt}
\tablecaption{Member galaxy properties and spectroscopic data used in this study. The following papers are referenced: [1] \cite{Hashimoto2023sep} [2] \cite{Venturi2024nov}; [3] \cite{Morishita2025may}; [4] \cite{Morishita2023apr}; [5] \cite{Witten2025jul}; [6] \cite{Naidu2024oct}. The magnification factor, $\mu$, was derived in Appendix~\ref{subsec:Lens_Model}.}\label{tab:spec_obs}
\tablehead{
\colhead{Object Name} &\colhead{RA} &\colhead{Dec} & \colhead{Program} & \colhead{Observation Mode} & \colhead{Disperser/Filter} & \colhead{$z$} & \colhead{$\mu$} & \colhead{Line Analyses}
}
\startdata
YD1& 3.60385 & -30.38189 &GO-1840 & NIRSpec/IFS &G395H/F290LP & $7.8778 \pm 0.0001$\textsuperscript{[1]} & 1.94 & YES \\ 
YD4 & 3.60381 & -30.38224 & GO-1840 & NIRSpec/IFS & G395H/F290LP    & $7.8742 \pm 0.0001$\textsuperscript{[1]} & 1.95 & YES \\
YD6 & 3.60403 & -30.38229 & GO-1893 & NIRSpec/IFS & PRISM/CLEAR & $7.880 \pm 0.001$\textsuperscript{[2]} & 1.95 & NO \\
YD7W & 3.60326 & -30.38224 & GO-1840 & NIRSpec/IFS & G395H/F290LP    & $7.8721 \pm 0.0001$\textsuperscript{[1]} & 1.96 & YES  \\ 
YD7E & 3.60346 & -30.38225 & GO-1840 & NIRSpec/IFS & G395H/F290LP & $7.93\pm0.04\tablenotemark{\rm{*}}$ & 1.96 & NO \\
s1& 3.60366 & -30.38193 & GO-1840 & NIRSpec/IFS & G395H/F290LP& $7.8732 \pm 0.0001$\textsuperscript{[1]} & 1.94 &  YES \\
ZD1 & 3.60358 & -30.38243 & GO-1840 & NIRSpec/IFS & G395H/F290LP& $7.9\pm0.2\tablenotemark{\rm{*}}$ & 1.96 & NO  \\
ZD3& 3.60642 & -30.38098 & GTO-4553 & NIRSpec/IFS & G395H/F290LP& $7.8808 \pm 0.0001$\textsuperscript{[3]} & 1.87 &  YES \\
ZD6& 3.60666 & -30.38088 & GTO-4553 & NIRSpec/IFS & G395H/F290LP& $7.8797 \pm 0.0001$\textsuperscript{[3]} & 1.87 & YES   \\
ZD12E& 3.60700 & -30.38069 & GTO-4553 & NIRSpec/IFS & G395H/F290LP&$7.8782 \pm 0.0001$\textsuperscript{[3]} & 1.86 &  YES  \\
ZD12W& 3.60695 & -30.38083 & GTO-4553 & NIRSpec/IFS & G395H/F290LP&$7.8759 \pm 0.0001$\textsuperscript{[3]} & 1.86 & YES   \\
ZD2& 3.60454 & -30.38043 & ERS-1324 & NIRSpec/MOS & G395H/F290LP& $7.8800 \pm 0.0001$\textsuperscript{[4]} & 1.88 & YES  \\ 
ZD4& 3.60525 & -30.38058 & GO-2561 & NIRSpec/MOS &PRISM/CLEAR& $7.883 \pm 0.001$\textsuperscript{[5]}  & 1.87 & YES \\ 
GLASSz8-2& 3.60134 & -30.37924 & GO-2561 & NIRSpec/MOS &PRISM/CLEAR&  $7.8831 \pm 0.0001$\textsuperscript{[4]} & 2.02 &YES \\
YD8& 3.59608 & -30.38583 & DD-2756 & NIRSpec/MOS &PRISM/CLEAR    & $7.8869 \pm 0.0001$\textsuperscript{[4]} & 2.76 &YES   \\
PC2E & 3.58792 & -30.37631 & UNCOVER & NIRSpec/MOS &PRISM/CLEAR & $7.8798 \pm 0.001$\textsuperscript{[5]} & 2.73&YES \\
PC5 & 3.58898 & -30.37865 & UNCOVER & NIRSpec/MOS &PRISM/CLEAR & $7.8709 \pm 0.001$\textsuperscript{[5]} & 2.71 & YES\\
ALT-41799 & 3.59513 & -30.38112 & ALT & NIRCam/WFSS & F356W & $7.883 \pm 0.005$\textsuperscript{[6]} & 2.60 & NO\\
PC1 & 3.60444 & -30.38026 & -- & -- & -- & $7.88\pm0.04\tablenotemark{\rm{*}}$ & 1.88 & NO \\ 
PC2W & 3.58778 & -30.37628 & -- & -- & -- & $7.93\pm0.03\tablenotemark{\rm{*}}$ & 2.74 & NO \\ 
PC3 & 3.61351 & -30.43474 & -- & -- & -- & $7.88\pm0.09\tablenotemark{\rm{*}}$ &1.53 &NO \\ 
PC4 & 3.61573 & -30.36839 & -- & -- & -- & $7.89\pm0.05\tablenotemark{\rm{*}}$ & 1.57 & NO \\ 
PC6 & 3.60013 & -30.38350 & -- & -- & -- & $7.91\pm0.09\tablenotemark{\rm{*}}$ & 2.14 & NO \\ 
\enddata
\tablenotetext{*}{Photometric redshift from \cite{Witten2025jul}.}
\tablecomments{In the NIRSpec data used in this study, the emission lines of YD6 could not be separated from those of YD4, and we therefore do not perform an emission-line analysis for YD6. We note, however, that \cite{Venturi2024nov} reported the detection of emission lines from this source and spectroscopically confirmed it at $z = 7.880$. ALT-41799 was spectroscopically identified with NIRCam/WFSS, but it was not observed with the filter covering [O{\sc iii}]5008, nor has it been observed with NIRSpec. We therefore exclude this source from the analyses based on emission-line fluxes in this study.}
\end{deluxetable*}

\section{Data} \label{sec:Data}
\subsection{NIRCam Imaging Data}\label{subsec:NIRCam_Imaging_Data}
NIRCam observations of the Abell 2744 field have been carried out by a series of JWST programs.
In this study, we used the publicly available UNCOVER DR3 data\footnote{\url{https://jwst-uncover.github.io/DR3.html\#Mosaics}} \citep{Suess2024nov, Bezanson2024oct}. We utilized imaging data in a total of 19 bands, consisting of eight JWST broad-band filters (F070W, F090W, F115W, F150W, F200W, F277W, F356W, and F444W) and eleven medium-band filters (F140M, F182M, F210M, F250M, F300M, F335M, F360M, F410M, F430M, F460M, and F480M). These observations were obtained through the following JWST programs: Medium Bands, Mega Science (GO-4111; PI: Suess; \citealt{Suess2024nov}), MAGNIF (GO-2883, PI: Sun; \citealt{Sun2023may}), ALT (GO-3516; PI: Naidu \& Matthee; \citealt{Naidu2024oct}), GO-3538 (PI: Iani), UNCOVER (GO-2561; PI: Labbe; \citealt{Bezanson2024oct}), GLASS-JWST (ERS-1324; PI: Treu; \citealt{Treu2022aug}), and DD-2767 (PI: Chen). 

The UNCOVER DR3 mosaic images have already been reduced; details of the data reduction procedures are provided in \cite{Suess2024nov} and \cite{Bezanson2024oct}. These data were processed using the \texttt{Grizli} pipeline \citep{Brammer2021jun, Brammer2022jun}. Within \texttt{Grizli}, an updated version of \texttt{snowblind}\footnote{\url{https://github.com/mpi-astronomy/snowblind}} was used to mask snowballs and bad pixels. The absolute astrometry of these mosaic images was calibrated using the NOAO Legacy Survey DR9 catalog \citep{Dey2019may}.

For each band, we derived an empirical point spread function (PSF) by stacking point sources. We then performed PSF homogenization using \texttt{PyPHER} \citep{Boucaud2016dec}. The mosaic images used in our analysis were matched to the F480M image, which has the broadest PSF among the filters adopted in this study. The $5\sigma$ limiting magnitudes of the imaging data after PSF homogenization, measured within a circular aperture of radius $0\farcs16$, range from $27.58\ (\rm{F480M})$ to $29.08\ (\rm{F356W})$ mag.

Because many of the galaxies studied here are located close to neighboring sources, curve-of-growth photometric measurements are susceptible to contamination and do not provide robust total flux. We therefore did not apply aperture corrections. Instead, for each source, we adopted a single elliptical aperture common to all NIRCam bands. The aperture size and shape were chosen to enclose the $1.5\sigma$ isophotal regions in the PSF-matched images across all bands, while avoiding contamination from nearby galaxies. The photometry of the member galaxies was then measured within these apertures using \texttt{photutils} \citep{Bradley2023may}.

The aperture sizes vary from galaxy to galaxy, with semi-major and semi-minor axes of $a = 0.2\text{--}0.6$ arcsec and $b = 0.14\text{--}0.46$ arcsec, respectively. The local sky background was measured for each member galaxy within an annulus centered on the source, with inner and outer radii of $1\farcs0$ and $2\farcs0$, respectively\footnote{To avoid contamination of the sky background estimate by signals from nearby sources, we applied sigma clipping to the annular region ($\sigma = 2.0$, \texttt{maxiters} = 5). The local sky background was then estimated by multiplying the mode within the annulus (in units of 10 nJy/pixel) by the aperture area of the source (in units of pixels), and this value was subtracted from the measured photometry.} (c.f., \citealt{Mawatari2026feb}). The photometric uncertainties were estimated as the standard deviation of fluxes measured from 3000 randomly placed apertures with the same size as the source aperture in source-free regions, after masking bright objects around the member galaxies. 
To account for residual systematic uncertainties in the NIRCam photometry from PSF matching, aperture definition, and background subtraction, we imposed a conservative minimum uncertainty floor of 5\%.
The measured photometry before correcting for gravitational lensing magnification is summarized in Tables~\ref{tab:Observed_measurements_table1}, \ref{tab:Observed_measurements_table2}, and \ref{tab:Observed_measurements_table3}. The correction for gravitational lensing magnification is described in Appendix~\ref{subsec:Lens_Model} (see Table~\ref{tab:spec_obs}).

\subsection{NIRSpec Spectroscopy Data}\label{subsec:NIRSpec_Spectroscopy_Data}
NIRSpec observations of the member galaxies in this protocluster have been carried out through several JWST programs. We make use of both Integral Field Spectroscopy (IFS) and Multi-Object Spectroscopy (MOS) data to investigate the properties of the member galaxies (YD1, YD4, YD7W, s1, ZD3, ZD6, ZD12E, ZD12W, ZD2, ZD4, GLASSz8-2, YD8, PC2E, and PC5). Details of the data used for each member galaxy are summarized in Table \ref{tab:spec_obs}.

\subsubsection{Integral Field Spectroscopy}\label{subsubsec:Integral_Field_Spectroscopy}
For the galaxies in ``the Quintet'' and ``the Chain'' (YD1, YD4, YD7W, s1, ZD3, ZD6, ZD12E, and ZD12W), we used NIRSpec IFS data from two JWST programs. The galaxies in ``the Quintet'' region (YD1, YD4, YD7W, and s1) were observed as part of the RIOJA project (JWST GO1 PID 1840; PIs: J. Álvarez-Márquez and T. Hashimoto; e.g., \citealt{Alvarez-Marquez2021mar, Hashimoto2023sep}). The galaxies in ``the Chain'' region (ZD3, ZD6, ZD12E, and ZD12W) were observed through GTO-4553 (PI: M. Stiavelli; \citealt{Morishita2025may}).

In both programs, the NIRSpec/IFS data were taken with the G395H/F290LP grating/filter configuration. This setup provides high spectral resolution, $R \sim 2700$, over the wavelength range $2.87$--$5.27$ $\mu$m.
We reduced the data following the standard procedures with the official JWST pipeline (ver. 1.19.0; \citealt{Bushouse2023jul}) under CRDS context \texttt{jwst\_1408.pmap}. 
To improve the data quality, we applied custom processing steps, including (1) the removal of cosmic rays and bad pixels through sigma clipping and (2) subtraction of the moving average background from the calibrated images (e.g., \citealt{Ubler2023sep, Perna2023nov, Marshall2023oct}). 
These steps are implemented in our custom codes \texttt{jwstgo1840}\footnote{\url{https://github.com/sugayu/jwstgo1840}}.
The final data cubes have a pixel scale of $0\farcs05$.
We used NIRSpec IFU observations of the A3V standard star 1808347 (2MASS J18083474+6927286) to evaluate the wavelength-dependent PSF of the IFS data. These data were obtained as part of the JWST commissioning program (PID 1128; PI: N. Lützgendorf) with the same filter--grating configuration as our science observations. We reduced the standard-star data using the same JWST pipeline version and CRDS context as those adopted for A2744-z7p9OD. We then performed PSF matching using convolution kernels generated with \texttt{pypher}.

We defined the apertures and extracted the 1D spectra as follows.
First, for galaxies in which the brightest detected emission line, [O{\sc iii}]5008 (c.f., Tables \ref{tab:Observed_measurements_table1}, \ref{tab:Observed_measurements_table2}, and \ref{tab:Observed_measurements_table3}), was detected, we placed elliptical apertures to approximately encompass the [O{\sc iii}]5008 line-emitting region of each galaxy. We then spatially integrated the signal within each aperture using \texttt{photutils} to obtain a preliminary 1D spectrum. Next, we fitted the [O{\sc iii}]5008 line with a single Gaussian function and constructed an integrated intensity map by integrating over the wavelength range spanning $\pm 1.5\times{\rm FWHM}$ around the line peak. Finally, for each galaxy, we defined an elliptical aperture that sufficiently enclosed the $1.5\sigma$ isophotal region in the [O{\sc iii}]5008 integrated intensity map. Here, $\sigma$ denotes the $1\sigma$ noise estimated from source-free sky pixels for each galaxy. ($1\sigma = 0.57$--$1.63 \times 10^{-20}\ \rm{erg\ s^{-1}\ cm^{-2}}$). By spatially integrating within this aperture, we extracted the 1D spectrum of each galaxy\footnote{\cite{Hashimoto2023sep} and \cite{Venturi2024nov} both measured the flux of YD1 using NIRSpec/IFS data, and the flux obtained in this study is consistent with their measurements within the uncertainties. We therefore regard the apertures adopted in this study as reasonable.}. 
% {\color{red}As an example, the spectrum of YD4 is shown in Figure \ref{YD4_spectrum}. The spectra of the other member galaxies are presented in Figure \ref{memgal_spectra}.}

Several studies have reported YD6 as a subcomponent of YD4 (e.g., \citealt{Hashimoto2023sep,Venturi2024nov,Fudamoto2025oct}). In our NIRSpec/IFS data, the emission from YD6 is too faint and blended with that of YD4 to be measured independently. We therefore do not perform a separate spectroscopic analysis for YD6, and the emission-line fluxes measured in this region are assigned to YD4.
YD7E is identified as a subcomponent of YD7W in the JWST/NIRCam imaging data, but no emission line is detected from YD7E in the NIRSpec/IFS data. We therefore exclude YD7E in the spectroscopic analysis.
We note that YD6 and YD7E are treated as separate components in the analysis based on the JWST/NIRCam imaging data.

In addition, no emission line was detected from ZD1 in ``the Quintet'' region (e.g., \citealt{Hashimoto2023sep,Venturi2024nov,Fudamoto2025oct}) at the wavelength corresponding to $z\sim7.88$. We exclude this source from the spectroscopic analysis, although it is included in the analysis based on the JWST/NIRCam imaging data.

\subsubsection{Multi-Object Spectroscopy}\label{subsubsec:Multi-Object_Spectroscopy}
We used the reduced spectroscopic data publicly available through the Dawn JWST Archive (DJA)\footnote{\url{https://dawn-cph.github.io/dja/index.html}} ver. 4.4 for MOS data. The DJA data were reduced using \texttt{msaexp} \citep{Brammer2022nov}. Here, we briefly summarize the DJA data reduction procedures (for details, see \citealt{deGraaff2025may,Heintz2024may}). 

The reduction starts from the public JWST exposure-level products from MAST and applies additional preprocessing, including corrections for 1/$f$ noise and residual bias offsets. The data are then processed through the standard NIRSpec calibration steps, including wavelength assignment, 2D spectral extraction, flat-field correction, path-loss and bar-shadow corrections, and photometric calibration. The calibrated 2D spectra are finally combined, and the 1D spectra are optimally extracted by taking an inverse-variance-weighted sum along the spatial direction.

\subsubsection{Line Flux Measurement}\label{subsubsec:Line_Flux_Measurement}
From the 1D spectra of each galaxy extracted from the NIRSpec IFS and MOS observations, we measured the emission-line fluxes of [O{\sc iii}]4960, 5008, H$\beta$, and [O{\sc ii}]3727, 3730. 
We first fitted the [O{\sc iii}]4960, 5008 lines, which are the brightest among the detected emission lines, with a double-Gaussian function and derived the redshift of each galaxy. 
By deconvolving the fitted profile with the instrumental spectral resolution of the disperser\footnote{\url{https://jwst-docs.stsci.edu/jwst-near-infrared-spectrograph/nirspec-instrumentation/nirspec-dispersers-and-filters\#gsc.tab=0}}, we derived the intrinsic FWHM of [O{\sc iii}]5008 for each galaxy. 

Next, we constructed the error spectrum for each galaxy. Based on the estimated redshift and intrinsic FWHM, we calculated the expected observed Gaussian width, $\sigma$, at the wavelength of each emission line, taking into account the wavelength dependence of the instrumental resolution. Using these widths, we masked the wavelength ranges within $\pm4\sigma$ around the bright emission lines\footnote{The masked lines are [O{\sc iii}]5008, 4960, H$\beta$, H$\gamma$, [Ne{\sc iii}]$\lambda$3870$\mathrm{\AA}$, and [O{\sc ii}]3727, 3730.}. After masking these regions, we constructed the error spectrum by calculating, for each spectral element, the standard deviation of the unmasked flux values within a sliding window of $\pm0.3\ \mu$m centered on that element.

We then measured the emission-line fluxes by applying Gaussian profile fitting to the spectra. To allow for the possibility that different emission lines may have different redshifts and line widths, we treated the redshift, line width, and amplitude as free parameters in the fitting. For [O{\sc iii}]4960, 5008, we assumed a common line width, fixed the wavelength separation, and adopted an amplitude ratio of [O{\sc iii}]5008 : [O{\sc iii}]4960 $= 2.98 : 1$ \citep{Storey2000mar}. H$\beta$ was fitted with a single Gaussian function. For [O{\sc ii}]3727, 3730, the G395H grating data provide sufficiently high spectral resolution ($R \sim 2700$) to resolve the two components, and we therefore fitted them with a double-Gaussian function, assuming a common line width and fixing their wavelength separation. In contrast, for the Prism data, the spectral resolution is insufficient to separate [O{\sc ii}]3727 and [O{\sc ii}]3730, and we therefore fitted them with a single Gaussian function. The fluxes were obtained by integrating the resulting Gaussian profiles over wavelength\footnote{We used the analytical integral of the Gaussian function.}. To evaluate the fitting uncertainties, we adopted a Monte Carlo (MC) approach. Specifically, we repeatedly performed Gaussian profile fitting on 1000 mock spectra generated by perturbing the observed spectrum according to the error spectrum. We then defined the 68\% confidence interval ($1\sigma$) of the flux from the scatter in the best-fit parameters obtained from these 1000 realizations. We adopted an integrated-flux SNR threshold of $\geq 3$ as the criterion for line detection. For non-detections, we estimated the $3\sigma$ upper limit by integrating the error spectrum over the range of $\pm 1.5\times {\rm FWHM}$ centered on the line center, where the FWHM was assumed to be that of [O{\sc iii}]5008 line, and multiplying the result by a factor of three. The resulting emission-line fluxes before the lensing magnification corrections are summarized in Tables \ref{tab:Observed_measurements_table1}, \ref{tab:Observed_measurements_table2}, and \ref{tab:Observed_measurements_table3}.

\section{SED Fitting} \label{sec:SED_Fitting}
\subsection{Modeling} \label{subsec:Modeling}
To investigate the star formation activity of the member galaxies, we performed SED fitting using the photometric measurements in a total of 19 bands derived in Section~\ref{subsec:NIRCam_Imaging_Data} and corrected for gravitational lensing magnification as described in Appendix~\ref{subsec:Lens_Model}.

We used the Bayesian Analysis of Galaxies for Physical Inference and Parameter EStimation code (\texttt{BAGPIPES}; \citealt{Carnall2018nov}) for the fitting\footnote{Following the \texttt{BAGPIPES} documentation, non-positive photometric fluxes were set to zero with uncertainties of $10^{99}$, effectively excluding them from the fit.}. 
For the Stellar Population Synthesis models, we adopted version 2.2.1 of the Binary Population and Spectral Synthesis code (\texttt{BPASS}; \citealt{Eldridge2017nov, Stanway2018sep}). 
We assumed a \cite{Kroupa2001apr} initial mass function with slopes of $-1.30$ for $0.1 < M/M_{\odot} < 0.5$ and $-2.35$ for $0.5 < M/M_{\odot} < 300$. In \texttt{BAGPIPES}, intergalactic medium (IGM) absorption is treated using the model of \cite{Inoue2014aug}, and \texttt{CLOUDY} (\citealt{Ferland1998jul, Ferland2017oct}) is implemented as the nebular emission model. 
For the dust attenuation curve, we adopted \cite{Calzetti2000apr}. For spectroscopically confirmed member galaxies, we fixed the redshift to the measured value for each source (Table \ref{tab:spec_obs}), while for member candidates without spectroscopic redshift confirmation, we fixed the redshift to $z=7.88$.

We assumed three parametric star formation histories (SFH) models (delayed-$\tau$, exponential, and constant) as well as a non-parametric SFH model. We confirmed that the derived stellar masses, SFRs, and dust attenuation values are broadly consistent among the different SFH assumptions. 
We therefore adopt the continuity non-parametric SFH model of \cite{Leja2019may}. For the non-parametric SFH model, we adopted the same time-bin setup as \citet{Fudamoto2025oct}: $0$--$10$, $20$--$50$, $50$--$100$, $100$--$150$, $150$--$300$, and $300$--$460$ Myr.
However, for PC3, for which photometric measurements are available in only nine filters, we adopt the results obtained with the delayed-$\tau$ model.

\subsection{Physical Properties from SED fitting} \label{subsec:Physical_Properties_from_SED_fitting}
SED fitting for our sample has been carried out in many previous studies (e.g., \citealt{Morishita2023apr, Morishita2025may, Hashimoto2023sep, Venturi2024nov, Witten2025jul}). In particular, \cite{Witten2025jul} performed SED fitting for 23 member galaxies in A2744-z7p9OD, including candidate members without spectroscopic confirmation, and derived physical properties such as stellar mass and SFR, but did not report the dust attenuation, $A_V$. As shown in section~\ref{sec:Line_Analyses}, $A_V$ is required to correct the emission-line fluxes for dust attenuation. We therefore performed our own SED fitting to derive $A_V$, stellar mass and SFRs.

\begin{figure}[t]
    \centering
    \includegraphics[width=\columnwidth,clip]{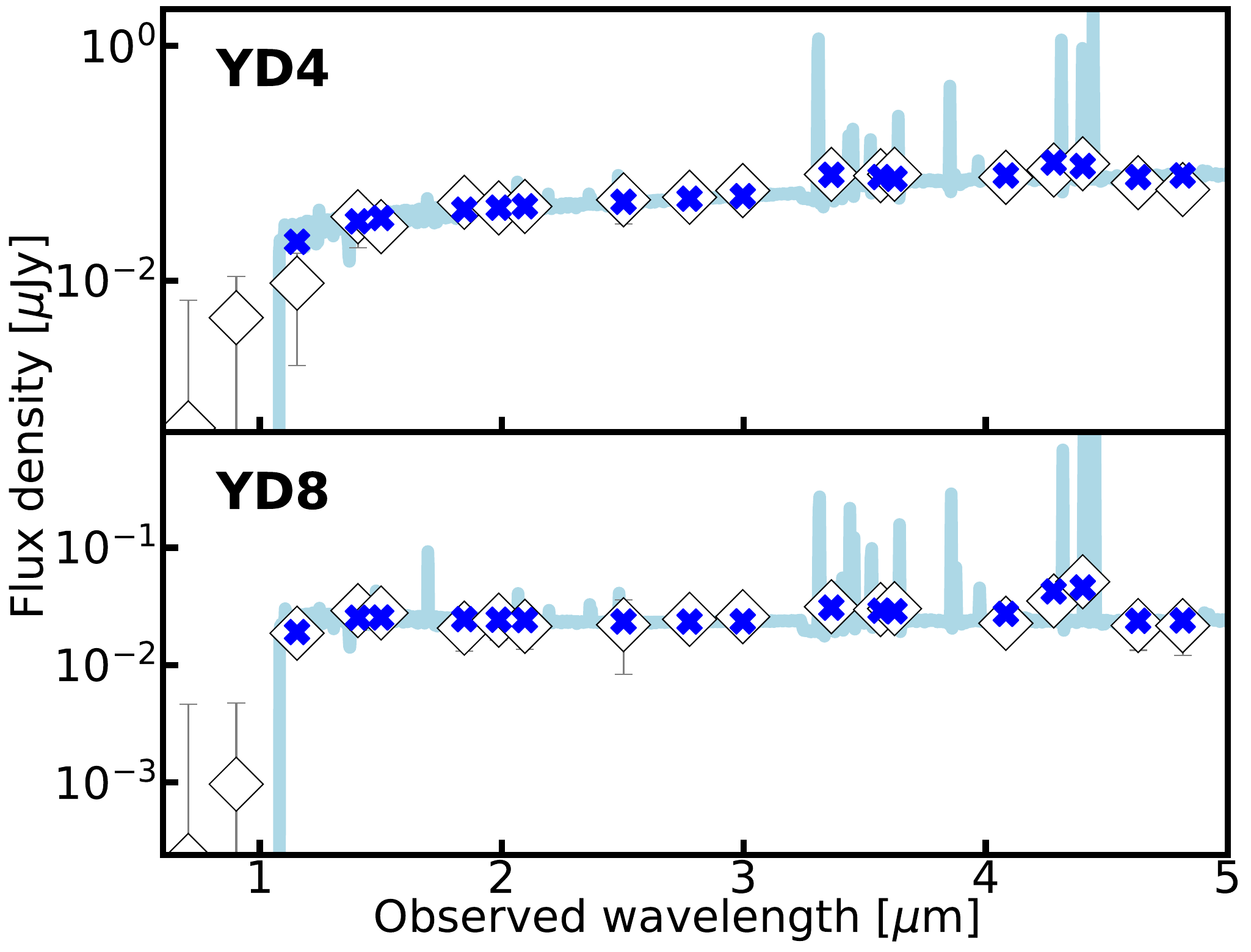}
    \caption{SEDs of YD4 (top) and YD8 (bottom). The diamonds represent the photometric measurements corrected for gravitational lensing magnification. The blue crosses and solid lines indicate the model photometry and best-fit SEDs, respectively. The error bars show the photometric uncertainties estimated from random-aperture measurements.}
    \label{fig:YD4_YD8_model_SED}
\end{figure}

Figure~\ref{fig:YD4_YD8_model_SED} shows the best-fit model SEDs for YD4 and YD8, while those for the other member galaxies are presented in Figure \ref{fig:memgal_model_SED}. Table \ref{tab:gal_properties} summarizes the physical properties of the individual member galaxies.

The stellar masses span a wide range, log $(M_\ast / M_{\odot}) \sim 7.5$ -- $9.2$, and the most massive galaxy is YD4, located in ``the Quintet'' region\footnote{The stellar masses of YD4 and YD7E are comparable. While \citet{Witten2025jul} identified YD7E as the most massive galaxy, YD4 is marginally more massive in our analysis, likely owing to differences in the assumptions and parameterizations adopted in the SED fitting. Nevertheless, the stellar-mass estimates for both galaxies agree with those of \citet{Witten2025jul} within the $1\sigma$ uncertainties.}.
Comparing the total stellar masses of the two dense galaxy groups, ``the Quintet'' and ``the Chain'', we obtain $\log (M_\ast/M_\odot)\sim9.8$ and $9.2$, respectively, indicating that ``the Quintet'' is approximately 4 times more massive than ``the Chain''. These results support the possibility that ``the Quintet'' is the most massive substructure within A2744-z7p9OD and suggest that the center of this protocluster may lie within the ``the Quintet'' region.

We derived SFRs on two different timescales, 10 Myr\footnote{Although recent SFRs on $\sim10$ Myr timescales can be estimated from Balmer recombination lines such as H$\alpha$ or H$\beta$ (e.g., \citealt{Kennicutt2012sep}), H$\beta$ is undetected for many galaxies in our sample, and the MOS spectra are affected by uncertain slit losses. We therefore adopt $\mathrm{SFR}_{10{\rm Myr}}$ derived from the SED fitting.} and 100 Myr, denoted as $\mathrm{SFR_{10Myr}}$ and $\mathrm{SFR_{100Myr}}$, respectively\footnote{These quantities were calculated by integrating the inferred SFH over the corresponding lookback-time interval and dividing the resulting stellar mass formed by 10 or 100 Myr. Thus, $\mathrm{SFR_{10Myr}}$ traces the most recent star formation activity, whereas $\mathrm{SFR_{100Myr}}$ reflects star formation activity over a longer timescale.}. 

The ratio of these two quantities, $\mathrm{SFR}_{10\mathrm{Myr}} / \mathrm{SFR}_{100\mathrm{Myr}}$, can thus be used to assess the recent burstiness of star formation in each galaxy. Specifically, values below unity indicate declining recent star formation activity, whereas values above unity imply enhanced recent star formation activity.

We find that $\mathrm{SFR}_{10\mathrm{Myr}} / \mathrm{SFR}_{100\mathrm{Myr}} \sim 0.4$--$9.2$, indicating a wide diversity in recent star formation activity among the galaxies. The specific star formation rate, defined as $\mathrm{sSFR_{10Myr}} \equiv \mathrm{SFR}_{10\mathrm{Myr}} / M_{\ast}$, also shows substantial scatter, with $\log (\mathrm{sSFR}_{10\ \mathrm{Myr}} / \mathrm{yr}^{-1}) \sim -8.8$ to $-7.0$. 

\begin{figure*}[t]
    \centering
    \includegraphics[width=0.99\textwidth,clip]{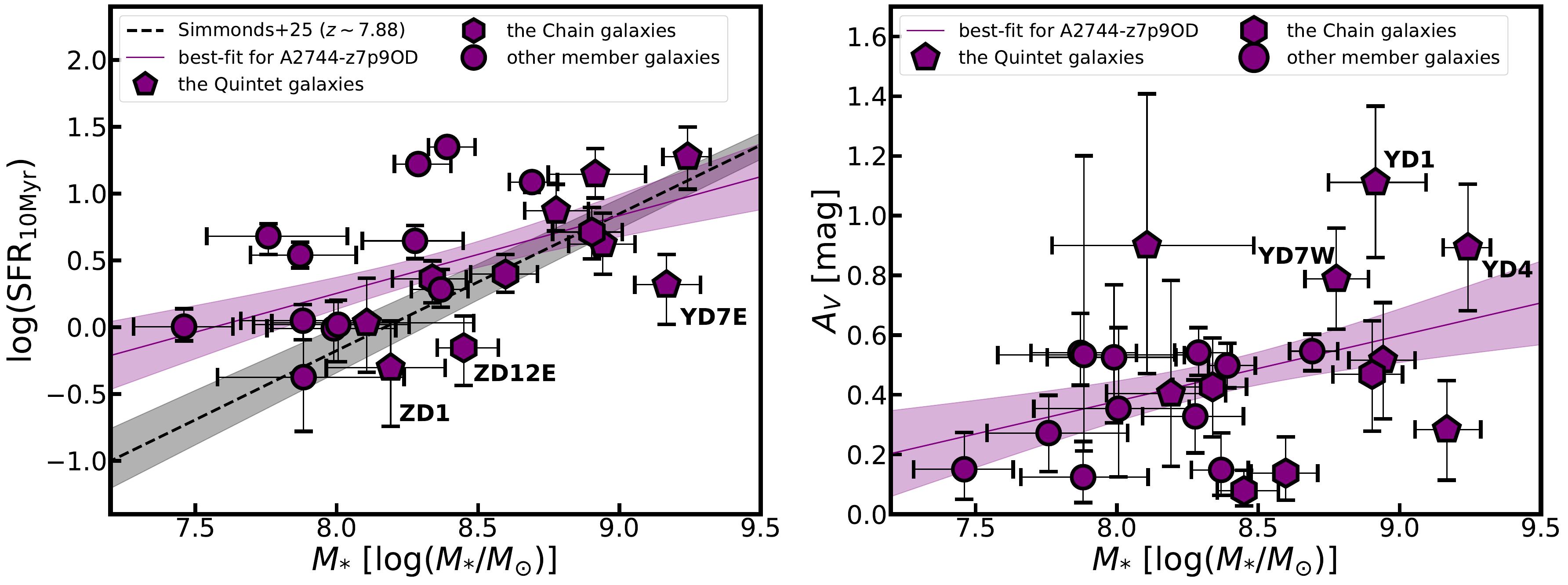}
    \caption{(Left panel) Stellar mass vs $\mathrm{SFR}_{10{\rm Myr}}$ diagram for the member galaxies of A2744-z7p9OD: pentagons for galaxies in ``the Quintet'', hexagons for those in ``the Chain'', and circles for the other members. The purple solid line and shaded region indicate the best-fit linear relation for member galaxies and its $1\sigma$ uncertainty, respectively. The black dashed line and shaded region indicate the typical relation and its $1\sigma$ scatter at $z=7.88$ from \citet{Simmonds2025dec}.
    (Right panel) Stellar mass vs $A_V$ diagram for the member galaxies. The purple solid line and shaded region indicate the best-fit linear relation and its $1\sigma$ uncertainty, respectively.
    }
    \label{fig:Mstar_SFR_Av}
\end{figure*}

The left panel of Figure~\ref{fig:Mstar_SFR_Av} shows the relation between stellar mass and $\mathrm{SFR}_{10{\rm Myr}}$ for the member galaxies. The purple solid line and shaded region indicate the best-fit linear relation and its 1$\sigma$ uncertainty for member galaxies, $\log(\mathrm{SFR}_{10\mathrm{Myr}}{/ M_{\odot}\ \mathrm{{yr}^{-1}}})=(0.6\pm0.2)\times\log(M_{*}/M_{\odot})-4.4\pm1.7$. For comparison, the black dashed line and shaded region indicate the typical star-forming main sequence and its 1$\sigma$ uncertainty at $z=7.88$ inferred from the redshift-dependent relation of \citet{Simmonds2025dec}\footnote{\citet{Simmonds2025dec} performed SED fitting based on a non-parametric SFH and derived SFRs averaged over the most recent 10 Myr, as in this study. To compare our results with the typical sSFR at the redshift of A2744-z7p9OD, we used the redshift-dependent main-sequence relation of \citet{Simmonds2025dec}, assuming $z=7.88$.}.
The relation for A2744-z7p9OD is broadly consistent with the typical relation at the same redshift. However, several low-mass member galaxies lie above the typical relation, suggesting enhanced recent star formation relative to typical galaxies with similar stellar masses at $z=7.88$.

From the black dashed line, the typical value is $\log (\mathrm{sSFR}_{10{\rm Myr}} / \mathrm{yr}^{-1}) \sim -8.15_{-0.14}^{+0.14}$. YD7E, ZD1, and ZD12E have lower central values of $\log (\mathrm{sSFR}_{10{\rm Myr}} / \mathrm{yr}^{-1}) = -8.8^{+0.3}_{-0.4}$, $-8.5^{+0.5}_{-0.5}$, and $-8.6^{+0.2}_{-0.4}$, respectively, although ZD1 is still consistent with the main sequence within the $1\sigma$ uncertainties. This suggests that some member galaxies, particularly YD7E and ZD12E, have relatively low recent star formation activity consistent with the findings of \cite{Witten2025jul} and \cite{Fudamoto2025oct}.  Overall, A2744-z7p9OD contains galaxies with a wide range of recent star formation activity, consistent with \citet{Witten2025jul}.

The right panel of Figure~\ref{fig:Mstar_SFR_Av} shows the relation between stellar mass and $A_{V}$ for the member galaxies. The purple solid line and shaded region indicate the best-fit linear relation and its 1$\sigma$ uncertainty for member galaxies, $A_V{/\mathrm{mag}}=(0.2\pm0.1)\times\log(M_\star/M_\odot)-1.4\pm1.0$.
Member galaxies with larger stellar masses tend to have higher $A_V$ values, consistent with previous findings for high-redshift galaxies (e.g., \citealt{Song2026may}).
The dust attenuation spans a range of $0.1$--$1.1$ mag, and we find that YD1, YD4, and YD7W, for which dust continuum detections by ALMA have been reported by \cite{Hashimoto2023sep}, \cite{Fudamoto2025oct} and \cite{Umehata2025nov}, exhibit particularly high dust attenuation among the member galaxies. This result is consistent with the redder UV continuum slopes reported for these galaxies by \citet{Hashimoto2023sep} and \citet{Witten2025jul}.

\begin{deluxetable*}{cccccccccccc}
% \digitalasset %提出時は復活？
\tablewidth{0pt}
\setlength{\tabcolsep}{2.6pt}
\tablecaption{Physical properties and emission-line ratios of the member galaxies. R23 and O32 are corrected for dust attenuation.}\label{tab:gal_properties}
\tablehead{ \colhead{} & \multicolumn{7}{c}{From SED fitting} 
     & \colhead{\texttt{galfit}} &\colhead{}&\multicolumn{2}{c}{Line ratio} \\ 
\colhead{} & \colhead{$\log M_\ast$} & \colhead{SFR$_{\rm{10 Myr}}$} & \colhead{SFR$_{\rm{100 Myr}}$} & \colhead{$\log\rm{sSFR}_{\rm{10 Myr}}$}& \colhead{$\frac{\rm{SFR_{\rm{10 Myr}}}}{\rm{SFR_{\rm{100 Myr}}}}$}  &  \colhead{$A_V$} & \colhead{$\chi^{2}_{\nu}$} &\colhead{$R_{\rm{e}}$}&\colhead{$\log\Sigma_{\rm{SFR}}$}
& \colhead{R23} & \colhead{O32}  \\ 
    &   \colhead{$M_\odot$} & \colhead{$M_\odot$ $\rm{yr}^{-1}$} & \colhead{$M_\odot$ $\rm{yr}^{-1}$} & \colhead{$\rm{yr}^{-1}$} & \colhead{}  & \colhead{mag}& \colhead{}  &\colhead{$10^{-1}$ kpc}&\colhead{$M_\odot{\rm yr}^{-1}{\rm kpc}^{-2}$}&\colhead{}&\colhead{}
}
\startdata
YD1&  ${8.9}_{-0.2}^{+0.2}$ &  ${14}_{-5}^{+8}$ &
        ${5.5}_{-1.9}^{+2.6}$ &  ${-7.8}_{-0.3}^{+0.2}$ &
        ${2.6}_{-1.0}^{+1.5}$  &
        ${1.1}_{-0.3}^{+0.3}$ & 1.4 & $7.5\pm0.4$  &${0.6}_{-0.2}^{+0.2}$& $>0.8$ & $0.4_{-0.1}^{+0.1}$\\
YD4&     ${9.2}_{-0.1}^{+0.1}$ &  ${19}_{-8}^{+13}$ &
        ${11}_{-3}^{+4}$ &  ${-8.0}_{-0.2}^{+0.2}$ &
        ${1.8}_{-0.6}^{+1.0}$ &  
        ${0.9}_{-0.2}^{+0.2}$ & 1.5 & $8.6\pm0.8$ &$0.6_{-0.3}^{+0.2}$& $>0.8$ & $0.0_{-0.1}^{+0.1}$\\
YD6&    ${8.9}_{-0.1}^{+0.1}$ &  ${4.2}_{-1.7}^{+3.0}$ &
        ${4.2}_{-1.5}^{+2.2}$ &  ${-8.3}_{-0.3}^{+0.2}$ &
        ${1.0}_{-0.5}^{+0.7}$ &  
        ${0.5}_{-0.2}^{+0.2}$ & 0.9 & $7.3\pm0.5$&$0.1_{-0.2}^{+0.3}$& - & - \\
YD7W&   ${8.8}_{-0.1}^{+0.1}$ &  ${7.5}_{-2.2}^{+4.3}$ &
        ${3.9}_{-1.1}^{+1.3}$ &  ${-7.9}_{-0.2}^{+0.2}$ &
        ${2.1}_{-0.7}^{+0.9}$ &  
        ${0.8}_{-0.2}^{+0.2}$ & 1.8 & $7.3\pm0.5$&$0.8_{-0.16}^{+0.2}$& $>0.9$ & $0.2_{-0.1}^{+0.1}$\\
YD7E&   ${9.2}_{-0.1}^{+0.1}$ &  ${2.1}_{-1.0}^{+1.4}$ &
        ${5.1}_{-2.1}^{+3.3}$ &  ${-8.8}_{-0.4}^{+0.3}$ &
        ${0.4}_{-0.2}^{+0.5}$ &  
        ${0.3}_{-0.2}^{+0.2}$& 0.9 & $6.0\pm0.2$ &$0.0_{-0.3}^{+0.2}$& - & - \\
s1&     ${8.1}_{-0.3}^{+0.4}$ &  ${1.1}_{-0.6}^{+1.3}$ &
        ${0.6}_{-0.3}^{+0.6}$ &  ${-8.0}_{-0.6}^{+0.4}$ &
        ${1.9}_{-1.1}^{+1.8}$ &  
        ${0.9}_{-0.4}^{+0.5}$& 2.6 &$5.7\pm0.4$&$-0.3_{-0.4}^{+0.3}$& $0.8_{-0.1}^{+0.1} $ & $ >-0.3$\\
ZD1&    ${8.2}_{-0.2}^{+0.2}$ &  ${0.5}_{-0.3}^{+0.6}$ &
        ${0.5}_{-0.2}^{+0.5}$ &  ${-8.5}_{-0.5}^{+0.5}$ &
        ${0.9}_{-0.6}^{+1.1}$ &  
        ${0.4}_{-0.2}^{+0.4}$& 1.0 &$4.5\pm0.7$&$-0.4_{-0.5}^{+0.4}$& - & - \\
ZD3&    ${8.6}_{-0.1}^{+0.1}$ &  ${2.5}_{-0.7}^{+1.0}$ &
        ${2.0}_{-0.5}^{+0.9}$ &  ${-8.2}_{-0.2}^{+0.2}$ &
        ${1.3}_{-0.5}^{+0.8}$ &  
        ${0.1}_{-0.1}^{+0.1}$& 0.8 &$4.2\pm0.2$&$0.4_{-0.1}^{+0.2}$& $0.9_{-0.1}^{+0.2}$ & $0.6_{-0.1}^{+0.2}$\\
ZD6&    ${8.9}_{-0.1}^{+0.1}$ &  ${5.2}_{-1.9}^{+2.7}$ &
        ${3.9}_{-1.4}^{+1.7}$ &  ${-8.2}_{-0.3}^{+0.3}$ &
        ${1.3}_{-0.5}^{+0.9}$ &  
        ${0.5}_{-0.2}^{+0.2}$&  1.5 &$5.2\pm0.5$&$0.5_{-0.2}^{+0.2}$& $1.1_{-0.1}^{+0.1} $ & $0.4_{-0.1}^{+0.1}$\\
ZD12E&    ${8.5}_{-0.1}^{+0.1}$ &  ${0.7}_{-0.3}^{+0.5}$ &
        ${1.2}_{-0.4}^{+0.5}$ &  ${-8.6}_{-0.4}^{+0.2}$ &
        ${0.6}_{-0.4}^{+0.6}$ &  
        ${0.1}_{-0.1}^{+0.1}$& 1.8 &$8.7\pm0.8$&$-0.8_{-0.3}^{+0.2}$& $0.4_{-0.1}^{+0.1}$ & $>0.4$\\
ZD12W&   ${8.3}_{-0.1}^{+0.1}$ &  ${2.3}_{-0.8}^{+0.8}$ &
        ${1.2}_{-0.4}^{+0.5}$ &  ${-8.0}_{-0.3}^{+0.2}$ &
        ${1.9}_{-0.7}^{+1.0}$ &  
        ${0.4}_{-0.2}^{+0.2}$& 1.7 &$2.6\pm0.4$&$0.8_{-0.2}^{+0.2}$& $0.8_{-0.1}^{+0.1}$ & $0.6_{-0.1}^{+0.1}$\\
ZD2&  ${8.4}_{-0.1}^{+0.1}$ &  ${22}_{-3}^{+3}$ &
        ${2.5}_{-0.3}^{+0.5}$ &  ${-7.0}_{-0.1}^{+0.1}$ &
        ${9.2}_{-1.6}^{+0.7}$ &  
        ${0.5}_{-0.1}^{+0.1}$& 3.1 &$2.6\pm0.1$&$1.7_{-0.1}^{+0.1}$& $1.0_{-0.1}^{+0.1}$ & $>1.0$\\
ZD4&    ${8.0}_{-0.2}^{+0.2}$ &  ${1.0}_{-0.4}^{+0.6}$ &
        ${0.5}_{-0.2}^{+0.4}$ &  ${-8.0}_{-0.5}^{+0.3}$ &
        ${1.9}_{-1.0}^{+1.5}$ &  
        ${0.5}_{-0.2}^{+0.2}$& 0.9 &$1.1\pm0.1$&$1.1_{-0.3}^{+0.2}$& $0.6_{-0.1}^{+0.1}$ &$>0.0$\\
GLASSz8-2&    ${8.3}_{-0.1}^{+0.1}$ & ${17}_{-2}^{+3}$ &
        ${2.0}_{-0.3}^{+0.5}$ & ${-7.1}_{-0.1}^{+0.1}$ &
        ${8.8}_{-1.9}^{+1.1}$ & 
        ${0.5}_{-0.1}^{+0.1}$& 2.1 &$1.5\pm0.1$&$2.1_{-0.1}^{+0.1}$& $1.1_{-0.1}^{+0.1}$ &$1.0_{-0.1}^{+0.2}$\\
YD8&    ${8.3}_{-0.2}^{+0.2}$ &  ${4.4}_{-1.2}^{+1.4}$ &
        ${1.4}_{-0.4}^{+0.5}$ &  ${-7.6}_{-0.3}^{+0.2}$ &
        ${3.3}_{-1.3}^{+1.7}$ &  
        ${0.3}_{-0.1}^{+0.1}$& 0.7&$4.4\pm0.1$&$0.6_{-0.1}^{+0.1}$&$ >1.3$&$>0.7$\\
PC2E&    ${7.9}_{-0.2}^{+0.2}$ &  ${3.5}_{-0.7}^{+0.9}$ &
        ${0.6}_{-0.2}^{+0.3}$ &  ${-7.3}_{-0.3}^{+0.2}$ &
        ${5.6}_{-1.9}^{+2.3}$ &  
        ${0.5}_{-0.1}^{+0.1}$& 1.4 &$2.7\pm0.3$&$0.9_{-0.1}^{+0.2}$& $1.1_{-0.1}^{+0.1}$ &$>0.6$\\
PC5&    ${7.5}_{-0.2}^{+0.2}$ &  ${1.0}_{-0.2}^{+0.4}$ &
        ${0.2}_{-0.1}^{+0.1}$ &  ${-7.5}_{-0.2}^{+0.3}$ &
        ${4.4}_{-1.5}^{+2.1}$ &  
        ${0.2}_{-0.1}^{+0.1}$& 2.0  &$1.0\pm0.1$&$1.2_{-0.1}^{+0.2}$&$0.6_{-0.1}^{+0.1}$&$ >0.7$\\
ALT-41799&    ${8.7}_{-0.1}^{+0.1}$ &  ${12}_{-2}^{+2}$ &
        ${3.6}_{-0.7}^{+0.9}$ &  ${-7.6}_{-0.1}^{+0.1}$ &
        ${3.4}_{-0.8}^{+0.9}$ &  
        ${0.6}_{-0.1}^{+0.1}$ & 4.9 &$5.0\pm0.1$&$0.9_{-0.1}^{+0.1}$&-& - \\
PC1 &   $8.0_{-0.3}^{+0.3}$  & $1.0_{-0.5}^{+0.6}$ &$0.5_{-0.2}^{+0.4}$
         & $-8.0_{-0.5}^{+0.4}$ &$1.9_{-1.0}^{+1.6}$
         &  $0.4_{-0.2}^{+0.3}$
         & 0.5 &$7.0\pm0.7$&$-0.5_{-0.3}^{+0.2}$&-& - \\
PC2W &     $8.4_{-0.1}^{+0.1}$  & $1.9_{-0.5}^{+0.8}$ &$1.4_{-0.4}^{+0.5}$
         & $-8.1_{-0.2}^{+0.2}$ &  $1.5_{-0.5}^{+0.8}$
         &$0.2_{-0.1}^{+0.1}$
         & 0.9 &$2.7\pm0.1$&$0.6_{-0.1}^{+0.2}$&-& - \\
PC3 &    $7.8_{-0.2}^{+0.3}$  & $4.8_{-1.3}^{+1.2}$ &$0.6_{-0.3}^{+0.6}$
        & $-7.0_{-0.3}^{+0.1}$ &$9.1_{-4.6}^{+1.0}$
         &  $0.3_{-0.1}^{+0.1}$
        & 2.4&$1.4\pm0.1$ &$1.6_{-0.1}^{+0.1}$&-& - \\
PC4 &     $7.9_{-0.2}^{+0.2}$  & $1.1_{-0.3}^{+0.4}$ &$0.5_{-0.2}^{+0.3}$
         & $-7.8_{-0.3}^{+0.3}$ &$2.3_{-1.1}^{+1.7}$
         &  $0.1_{-0.1}^{+0.1}$
         & 0.7&$1.6\pm0.2$&$0.8_{-0.2}^{+0.2}$&-& - \\
PC6 &    $7.9_{-0.3}^{+0.4}$  & $0.4_{-0.3}^{+0.7}$ &$0.3_{-0.2}^{+0.4}$
         & $-8.3_{-0.5}^{+0.5}$ &$1.3_{-0.9}^{+1.2}$
         &  $0.5_{-0.3}^{+0.7}$
         & 0.8 &$2.4\pm0.5$&$0.1_{-0.4}^{+0.4}$&-& - \\
\enddata
\end{deluxetable*}

\section{Galaxy sizes}\label{sec:Galaxy_sizes}

We measured the rest-frame ultraviolet (UV) sizes of the member galaxies using \texttt{galfit} \citep{Peng2002jul,Peng2010jun}. We used the NIRCam F200W image before PSF matching, which corresponds to rest-frame $\simeq2250$~\AA\ at $z=7.88$. For this analysis, we used the F200W science image, weight map, and exposure-time map from DJA\footnote{The UNCOVER data release used for the photometric measurements does not include exposure-time maps. We therefore used the DJA data products for this analysis.}, so that the image products used to construct the error map were taken from the same reduction.

The error map was constructed following the method of \citet{Zhang2026feb}, combining the inverse-variance term from the weight map with the source Poisson-noise term estimated from the science image and exposure-time map. For each member galaxy, we fitted a single-component Sérsic profile convolved with the F200W PSF. The free parameters were the centroid position, total magnitude, effective radius, Sérsic index, axis ratio, and position angle. We also included a sky component in the fit, and, when nearby sources could affect the target profile, fitted them simultaneously as additional components. To correct for the spatial magnification due to gravitational lensing, we divided the effective radius obtained from the fitting by $\sqrt{\mu}$, assuming that the linear magnification factor is $\sqrt{\mu}$.

\begin{figure*}[t]
    \centering
    \includegraphics[width=0.99\textwidth,clip]{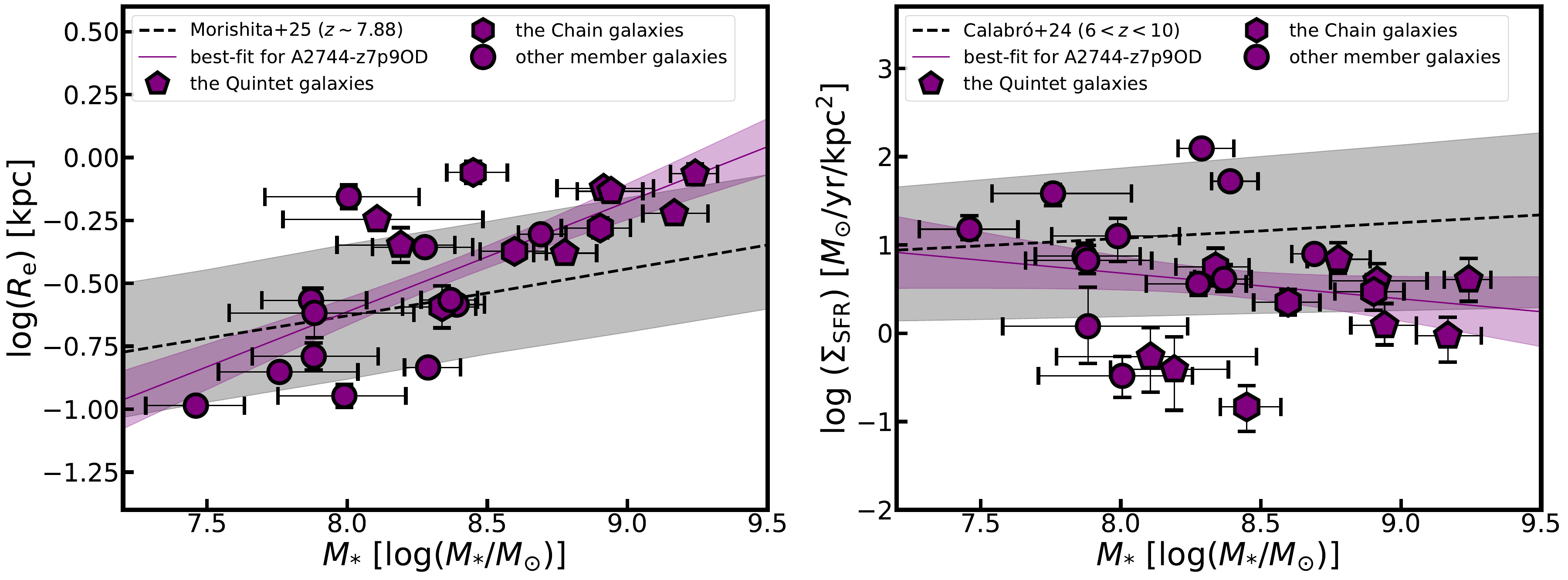}
    \caption{(Left panel) Stellar mass vs $R_{\mathrm{e}}$ diagram for the member galaxies of A2744-z7p9OD. The symbols are the same as in Figure~\ref{fig:Mstar_SFR_Av}. The purple solid line and shaded region indicate the best-fit linear relation for member galaxies and its $1\sigma$ uncertainty, respectively. The black dashed line and shaded region indicate the typical relation and its $1\sigma$ scatter at $z=7.88$ from \cite{Morishita2025may}.
    (Right panel) Stellar mass vs $\Sigma_{\rm{SFR}}$ for the member galaxies. The purple solid line and shaded region indicate the best-fit linear relation and its $1\sigma$ uncertainty, respectively. The black dashed line and shaded region indicate the typical relation and its $1\sigma$ scatter at $z=7.88$ from \cite{Calabro2024oct}.
    }
    \label{fig:Mstar_Re_SIGMASFR}
\end{figure*}

The resulting effective radii ($R_{\rm{e}}$) are listed in Table~\ref{tab:gal_properties}.
The fitted effective radii measured in the image plane satisfy $R_{\rm e} > {\rm FWHM\ PSF}/2$ for all member galaxies, where ${\rm FWHM\ PSF}\simeq0.064''$. Therefore, their effective diameters, $2R_{\rm e}$, are larger than the PSF FWHM. We thus regard these galaxies as at least marginally resolved (c.f. \citealt{Chen2022aug}).\footnote{\citet{Chen2022aug} suggest that \texttt{galfit} can reliably measure effective radii down to approximately one-third of the PSF FWHM, with uncertainties within 20\%.} Several studies of high-redshift galaxies use the semi-major-axis effective radius rather than the circularized effective radius (e.g., \citealt{Morishita2024mar, Calabro2024oct}). To facilitate comparison with these studies, we also adopt the semi-major-axis effective radius.  Derived effective radii span the range $ -1<\log(R_{\rm e}/\rm{kpc})<0$. The left panel of Figure~\ref{fig:Mstar_Re_SIGMASFR} shows the relation between effective radius and stellar mass. The purple solid line and shaded region indicate the best-fit linear relation and its 1$\sigma$ uncertainty for the member galaxies, $\log(R_{\rm e}{/{\rm kpc}})=(0.4\pm0.1)\times\log(M_{*}/M_{\odot})-4.1\pm0.8$. Compared with the typical relation at similar redshifts shown by the black dashed line \citep{Morishita2024mar}, $R_{\rm{e}}$ values are broadly consistent within the scatter.

We calculated the star-formation rate surface density, $\Sigma_{\rm SFR}$, using $\mathrm{SFR_{10Myr}}$ and $R_{\rm{e}}$. Assuming that half of the total star formation occurs within the effective radius, we defined the star formation rate surface density as
\begin{equation}
\Sigma_{\mathrm{SFR}} =  \mathrm{SFR}_{\mathrm{10Myr}}/ 2\pi R_{\mathrm{e}}^2 \label{equ:SIGMASFR} 
\end{equation}
where $\Sigma_{\mathrm{SFR}}$ is expressed in units of $M_\odot\ {\rm yr}^{-1}\ {\rm kpc}^{-2}$.

The derived $\Sigma_{\rm SFR}$ values are listed in Table~\ref{tab:gal_properties}. They span the range $-0.8 < \log(\Sigma_{\rm SFR}{/M_{\odot}\ {\mathrm{yr}^{-1}}\ \mathrm{kpc}^{-2})} < 2.1$. The right panel of Figure~\ref{fig:Mstar_Re_SIGMASFR} shows the relation between stellar mass and $\Sigma_{\rm SFR}$. The purple solid line and shaded region indicate the best-fit linear relation and its 1$\sigma$ uncertainty for the member galaxies, $\log(\Sigma_{\rm SFR}{/M_{\odot}\ {\mathrm{yr}^{-1}}\ {\mathrm{kpc^{-2}}}})=(-0.3\pm0.3)\times\log(M_{*}/M_{\odot})+3.0\pm2.7$. Compared with the $6<z<10$ relation of \citet{Calabro2024oct}, several member galaxies have lower $\Sigma_{\rm SFR}$, but most are broadly consistent with the relation within the uncertainties.

\begin{figure*}[t]
    \centering
    \includegraphics[width=0.97\textwidth,clip]{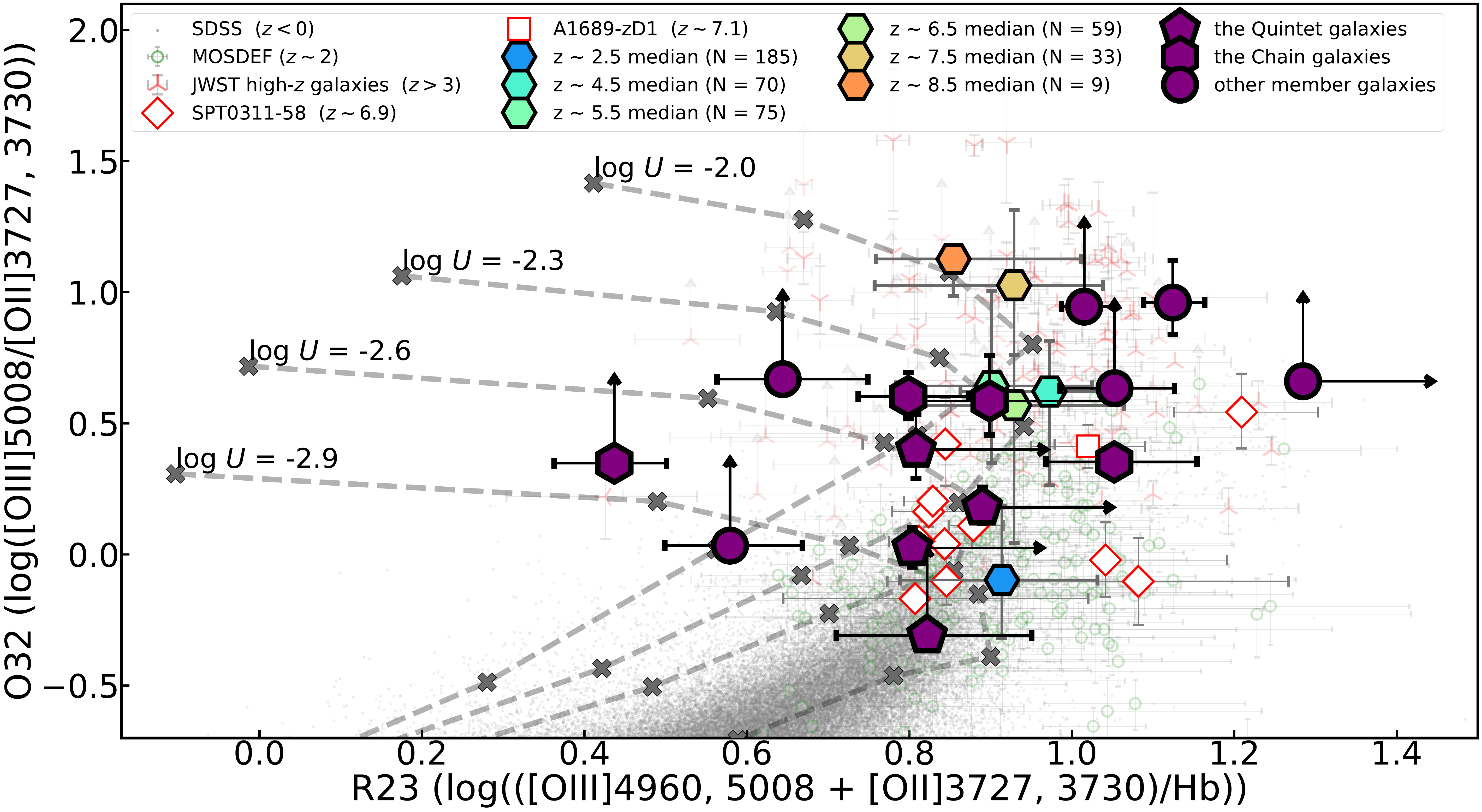}
    \caption{R23 vs O32 diagram. The member galaxies of A2744-z7p9OD ($z \sim 7.88$) are shown as purple symbols: pentagons for galaxies in ``the Quintet'', hexagons for those in ``the Chain'', and circles for the other members.  
    For comparison, individual galaxies at $z>3$ observed with JWST are shown as red symbols (\citealt{Cameron2023sep}; \citealt{Nakajima2023dec}; \citealt{Morishita2024aug}; \citealt{Arribas2024aug}; C. Blanco-Prieto et al., submitted). For galaxies included in both \citet{Cameron2023sep} and \citet{Morishita2024aug}, we adopt the measurements from \citet{Morishita2024aug}. Galaxies at $z\sim2$ from the MOSDEF survey are shown as green circles (\citealt{Kriek2015jun}; \citealt{Reddy2015jun}), and local galaxies at $z<0.1$ from SDSS are shown as gray points (\citealt{Aihara2011apr}). The median values and uncertainties in each redshift bin are shown as hexagons and the number of galaxies used to calculate each median is indicated in the caption. The gray dashed curves indicate the photoionization models of \citet{Kewley2002sep}; different curves correspond to different ionization parameters. From the upper left, the crosses along each curve indicate metallicities of $Z = 0.05$, 0.1, 0.2, 0.5, 1.0, 1.5, 2, and $3\,Z_{\odot}$.}
    \label{fig:R23_O32}
\end{figure*}

\section{Emission-Line Analyses} \label{sec:Line_Analyses}
\subsection{Correction For Dust Attenuation}\label{subsec:The_Correction_For Dust_Attenuation}
We correct the emission-line fluxes measured in Section~\ref{subsubsec:Line_Flux_Measurement} for dust attenuation. The dust attenuation toward nebular emission is often estimated from Balmer line ratios (e.g., H$\alpha$/H$\beta$). However, for the galaxies in our sample, H$\alpha$ falls outside the NIRSpec wavelength coverage, while H$\beta$ and H$\gamma$ are undetected in many cases. We therefore correct for dust attenuation using the stellar $A_V$ values derived from the SED fitting.

First, we convert the stellar $A_V$ into the nebular $A_V$ using the conversion factor $f = E(B-V)_{\rm star}/E(B-V)_{\rm neb}$. This factor remains uncertain for galaxies in the early Universe \citep{Song2026may}. In the absence of Balmer-decrement measurements, $f=1$ is often adopted for high-redshift galaxies (e.g., \citealt{Asada2024feb} for galaxies at $z \sim 6$), whereas \citet{Tsujita2026feb} recently reported $f = 0.51_{-0.03}^{+0.04}$ for main-sequence galaxies at $z \sim 4.4$–$5.7$.
In this study, we adopt $f=0.83$ as our fiducial value, following the empirical relation derived by \citet{Kashino2013nov} for a statistical sample of star-forming galaxies at $1.4 < z < 1.7$. Using this factor, we convert the stellar $A_V$ into the nebular $A_V$ and correct the emission-line fluxes using the \citet{Calzetti2000apr} attenuation curve, consistent with the SED fitting procedure.

\subsection{Line Ratios} \label{subsec:Line_Ratios}
We investigate the properties of the ionized gas in the member galaxies of A2744-z7p9OD through emission-line ratio diagnostics, using the dust-corrected fluxes of [O{\sc ii}]3727, 3730, H$\beta$, and [O{\sc iii}]5008. We focus on the following two line ratios:
\begin{equation}
\rm{R23} = \log{(([\mathrm{O}\text{\textsc{iii}}]5008,4960 + [\mathrm{O}\text{\textsc{ii}}]3727, 3730) / H\beta)} \label{equ:R23} 
\end{equation}
\begin{equation}
\rm{O32} = \log{([\mathrm{O}\text{\textsc{iii}}]5008 / [\mathrm{O}\text{\textsc{ii}}]3727, 3730)} \label{equ:O32}      
\end{equation}

R23, defined by Equation \ref{equ:R23}, has been widely used in previous studies as an emission-line ratio tracing the gas-phase metallicity (e.g., \citealt{Laseter2024jan, Nakajima2023dec, Morishita2024aug}). O32, defined by Equation \ref{equ:O32}, is an indicator of the ionization state of the gas and is primarily used as a tracer of the ionization parameter (e.g., \citealt{Nakajima2014jul, Kewley2019aug}). Since O32 also depends on metallicity, the combination of R23 and O32 enables a more robust characterization of the ionized-gas conditions in galaxies.

Table \ref{tab:gal_properties} summarizes the dust-corrected R23 and O32 values for each galaxy. Figure \ref{fig:R23_O32} shows the relation between the line ratios R23 and O32.
The member galaxies of A2744-z7p9OD are shown as purple symbols. For comparison, individual galaxies at $z > 3$ observed with JWST are plotted as red symbols (\citealt{Cameron2023sep,Nakajima2023dec,Morishita2024aug,Arribas2024aug}; C. Blanco-Prieto et al. submitted). 

To compare with lower-redshift galaxies, we show a sample at $z\sim2$ from the MOSDEF survey public release emission-line catalog\footnote{\url{https://mosdef.astro.berkeley.edu/for-scientists/data-releases/}} as green circles \citep{Kriek2015jun, Reddy2015jun}. 
To ensure reliability, we selected galaxies with [O{\sc ii}]3727, 3730, H$\beta$, [O{\sc iii}]5008, and H$\alpha$ detected at SNR $> 3$. For comparison with local galaxies, we also plot a sample at $z < 0.1$ from the SDSS MPA-JHU catalog\footnote{\url{https://www.sdss3.org/dr10/spectro/galaxy_mpajhu.php}} as gray points \citep{Aihara2011apr}. For this catalog, we used only galaxies whose spectra were flagged as ``reliable'' and selected those with [O{\sc ii}]3727, 3730, H$\beta$, [O{\sc iii}]5008, and H$\alpha$ detected at SNR $> 10$.  
To ensure reliable Balmer-decrement-based dust corrections, we applied the criterion H$\alpha$/H$\beta > 2.84$ to the literature samples, so that the observed Balmer decrements are physically consistent with the Case B recombination value.
The median value and its uncertainty in each redshift bin are indicated by hexagons, while the dashed curves show the photoionization models of \cite{Kewley2002sep}. In these models, the gas pressure is fixed at $P/k = 10^5\ \rm{cm}^{-3}$, while the metallicity and ionization parameter are allowed to vary.

The member galaxies of A2744-z7p9OD exhibit systematically lower O32 values on the R23--O32 plane than the typical values of galaxies at similar redshifts, $z \sim 7.5$ and $z \sim 8.5$ (O32 $\sim 1.0$ and $1.1$, respectively). Within A2744-z7p9OD, the lowest O32 values are found in ``the Quintet'' region.
In particular, YD4 and YD7W have O32 values of $\sim 0.0^{+0.1}_{-0.1}$ and $0.2^{+0.1}_{-0.1}$, comparable to the median value of MOSDEF galaxies at $z \sim 2$ (O32 $\sim -0.1$). Therefore, the protocluster core of A2744-z7p9OD is characterized by an ionization state as low as that of galaxies at later cosmic epochs.
A similarly low-ionization state has also been reported in the core of SPT0311-58, a protocluster at $z\sim6.9$, where the member galaxies show O32 values of $-0.2$ to $0.6$ \citep{Arribas2024aug}.
Taken together, these results demonstrate that galaxies in overdense environments, such as protocluster core regions, can exhibit lower ionization states than typical galaxies at similar redshifts during the EoR.

In addition, the member galaxies of A2744-z7p9OD show a wide range of R23, with $\Delta \mathrm{R23} \gtrsim 0.7$, which is larger than the measurement uncertainties. If interpreted using the calibration of \citet{Nakajima2022sep}, this range corresponds to $\Delta(12+\log({\rm O/H})) \gtrsim 0.9$. This suggests that significant galaxy-to-galaxy variations in gas-phase metallicity may already exist within A2744-z7p9OD. Similar variations have also been reported in the $z\sim6.9$ protocluster SPT0311-58 \citep{Arribas2024aug}. These results may indicate that chemical enrichment proceeds inhomogeneously among member galaxies in protoclusters even in the early Universe.

We also examined how the assumed conversion factor $f$ used in the dust attenuation correction affects the corrected line ratios.
For R23, adopting $f=0.51$ \citep{Tsujita2026feb} or $f=1$ instead of our fiducial value of $f=0.83$ changes the line ratios only within the uncertainties. 
For O32, the values obtained with $f=1$ are also consistent with the fiducial values within the uncertainties, while adopting $f=0.51$ systematically decreases O32 by $\sim 0.1$ dex. 
Therefore, the low O32 values of the protocluster-core galaxies are not caused by the particular choice of $f$.

\section{Protocluster Structure} \label{sec:Protocluster_Structure}
To quantitatively assess the impact of environment on galaxy evolution, it is essential to define environmental indicators based on the spatial distribution of protocluster member galaxies and to examine their relations with galaxy physical properties. We use the source-plane positions corrected for gravitational lensing and quantify the environment of each member galaxy using two structural parameters. Although it would be ideal to describe the galaxy distribution within the protocluster in full three dimensions, this is difficult in practice because the observed redshift contains contributions not only from the Hubble flow but also from peculiar motions within the protocluster, making it impossible to uniquely reconstruct the true line-of-sight position of each galaxy. 

We therefore evaluate the environment using the 2D projected distribution on the sky. Such analyses have been widely adopted in studies of protoclusters. For example, \cite{Perez-Martinez2023jan} introduced the local-density parameter $\Sigma_{3}$, defined from the third-nearest neighbor, for the Spiderweb protocluster and investigated its relation to galaxy physical properties. In addition, \cite{Champagne2025mar} examined environmental effects on galaxy evolution in overdensities at $z \sim 5.2, 6.2,$ and $ 6.6$ based on their 2D spatial distributions with JWST data. Following these previous studies, we parameterize the environment using their projected 2D distribution on the sky.

We first define $D_{\rm{nei}}$, the projected distance from each member galaxy to its nearest neighboring galaxy. This parameter can be interpreted as a local environmental indicator that reflects the local density, and is useful for evaluating whether close galaxy--galaxy interactions, including mergers, may affect galaxy properties.

Next, we introduce $D_{\rm YD4}$, the projected distance from YD4, as a global environmental indicator that traces the distance from the protocluster center. In a similar analysis of a protocluster at $z\sim3.70$, \citet{Toshikawa2025mar} adopted the mean position of the member galaxies as the protocluster center. However, such a definition can be affected by the incompleteness of the member-galaxy sample, especially if dusty starburst galaxies or quiescent galaxies are missed.
For A2744-z7p9OD, YD4 is the most massive member galaxy and is located in ``the Quintet'', a compact core region. A [C{\sc ii}] $158\,\mu$m survey of A2744-z7p9OD, which covers all member galaxies except for PC3 and PC4 (ALMA \#2023.1.01362.S), also indicates that the center of the gas and dust distribution is located in ``the Quintet'' region (Inui et al. in preparation). This supports the interpretation that the center of A2744-z7p9OD is associated with ``the Quintet''. To further support the use of YD4 as the central reference point, we calculated the stellar-mass centroid from the positions and stellar masses of the member galaxies and found that it is located only $\sim6.7\ {\rm pkpc}$ from YD4. We therefore adopt $D_{\rm YD4}$ as a practical proxy for the distance from the protocluster center.

\begin{figure}[t]
    \centering
    \includegraphics[width=1\columnwidth,clip]{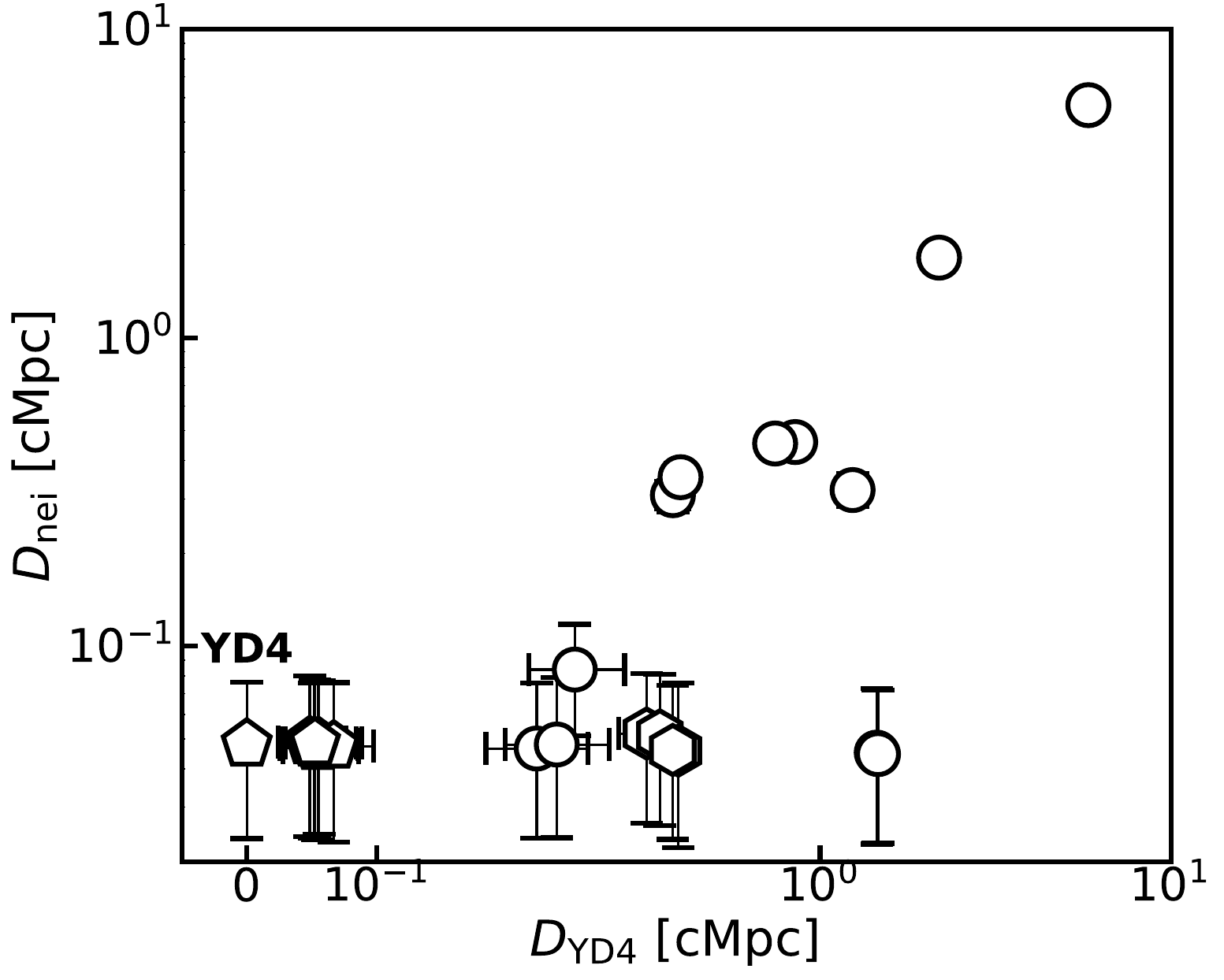}
    \caption{Plot of $D_{\rm{YD4}}$ vs $D_{\rm{nei}}$ for the member galaxies of A2744-z7p9OD: pentagons for galaxies in ``the Quintet'', hexagons for those in ``the Chain'', and circles for the other members. $D_{\rm YD4}$ is the projected distance from YD4, and $D_{\rm{nei}}$ is the projected distance from each member galaxy to its nearest neighboring galaxy.}
    \label{fig:structure_parameters}
\end{figure}

Figure \ref{fig:structure_parameters} shows the relation between $D_{\rm{YD4}}$ and $D_{\rm{nei}}$. The uncertainties were estimated from the lens-model uncertainty (RMS $\sim 0\farcs66$; \citealt{Furtak2023aug}). This figure suggests that galaxies located closer to the protocluster center tend to have nearby companions, whereas galaxies at larger distances tend to be more isolated. This relation implies that A2744-z7p9OD has a centrally concentrated structure.

\section{Discussion}\label{sec:Discussion}

\subsection{Overdensity Environment and Galaxy Properties} \label{subsec:Overdensity_Environment_and_Galaxy_Properties}

We have analyzed the stellar population properties (Section~\ref{subsec:Physical_Properties_from_SED_fitting}), galaxy sizes (Section~\ref{sec:Galaxy_sizes}), emission-line ratios (Section~\ref{subsec:Line_Ratios}), and structural parameters (Section~\ref{sec:Protocluster_Structure}) of the member galaxies in A2744-z7p9OD. By combining these measurements, we examine whether the galaxy properties are more strongly related to the global distance from the protocluster center, $D_{\rm YD4}$, or to the local environmental indicator, $D_{\rm nei}$. This comparison allows us to test whether the member-galaxy properties are primarily associated with the global protocluster environment or with local environment.

However, the spatial coverage of our dataset is limited compared with previous studies of protoclusters at lower redshifts. In particular, the projected area analyzed in this study is about a factor of $\sim  4$ smaller than that covered by \citet{Toshikawa2025mar} in comoving units. Because of this difference, a fair comparison of the absolute values of the structural parameters is not straightforward. We therefore focus on A2744-z7p9OD itself and examine which structural parameter better explains the properties of the member galaxies, based on the presence and strength of correlations with their physical properties.

\begin{figure*}[t]
    \centering
    \includegraphics[width=0.77\textwidth,clip]{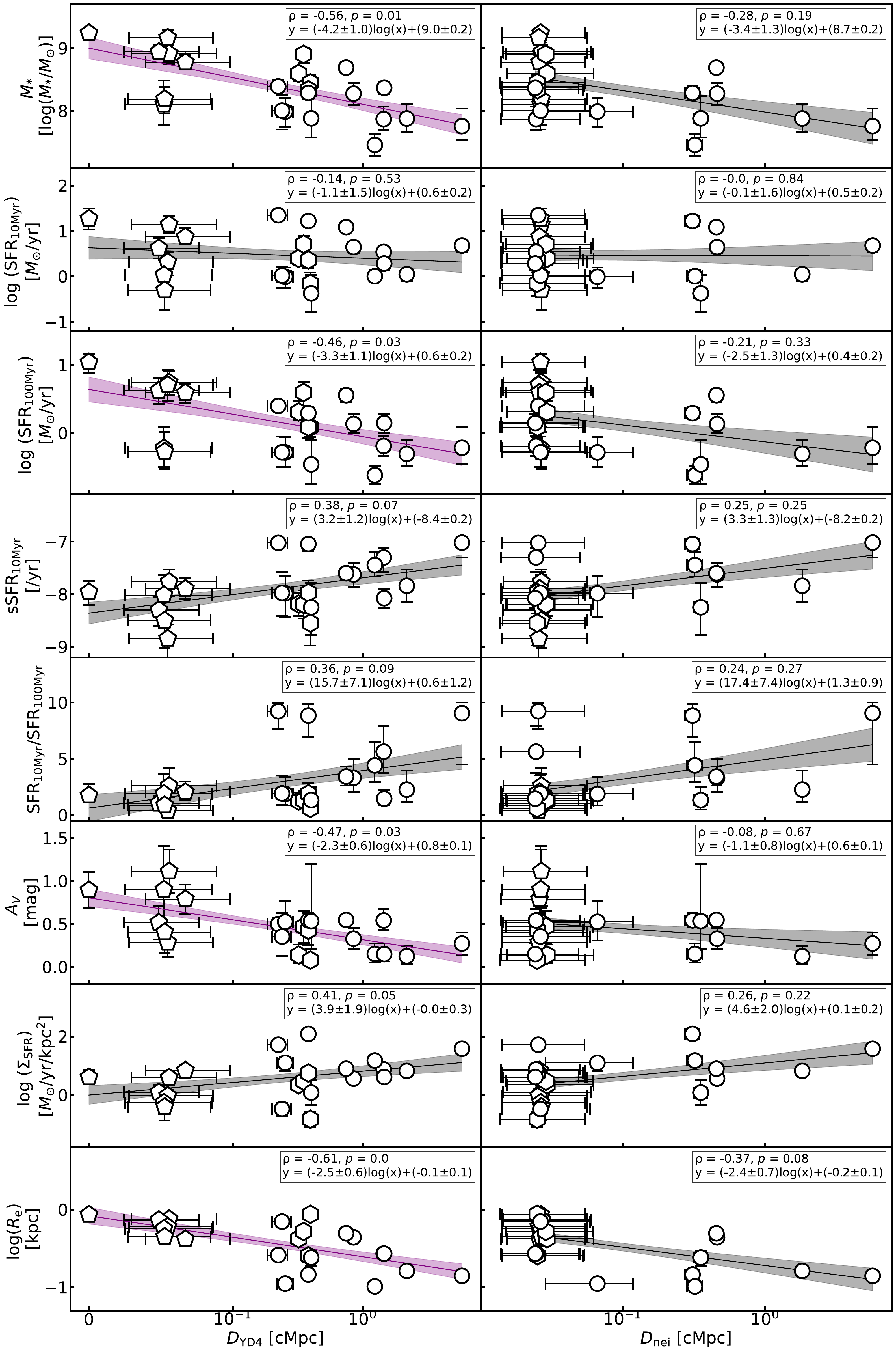}
    \caption{Stellar population properties of the member galaxies as a function of the two structural parameters. The columns correspond to $D_{\rm YD4}$ and $D_{\rm nei}$ from left to right, while the rows correspond to $M_\ast$, $\log\mathrm{SFR}_{10{\rm Myr}}$, $\log\mathrm{SFR}_{100{\rm Myr}}$, $\mathrm{sSFR}_{10{\rm Myr}}$, $\mathrm{SFR}_{10{\rm Myr}}/\mathrm{SFR}_{100{\rm Myr}}$, $A_V$, $\log\Sigma_{\rm{SFR}}$, and $\log R_{\rm{e}}$ from top to bottom. The Spearman rank correlation coefficient $\rho$, and the corresponding $p$-value are shown in the upper right corner of each panel. The solid lines show the best-fit linear relations, and the shaded regions indicate the corresponding $1\sigma$ uncertainties. Purple lines indicate statistically significant correlations ($p<0.05$), while gray lines indicate non-significant correlations.}
    \label{fig:distance_SEDproperties}
\end{figure*}

Figure~\ref{fig:distance_SEDproperties} shows the stellar population properties of the member galaxies as a function of the two structural parameters. To quantitatively investigate which structural parameter is more strongly associated with the star formation activity of the galaxies, we calculated the Spearman rank correlation coefficients. The correlation coefficient $\rho$, and the corresponding $p$-value are shown in the upper right corner of each panel. The equation of the best-fit linear relation is also shown in the same location. Correlations with $p < 0.05$ are considered statistically significant.

$M_*$ shows a significant negative correlation with $D_{\rm YD4}$, with $\rho=-0.56$ and $p=0.01$. $\mathrm{SFR}_{100{\rm Myr}}$ also shows a negative correlation with $D_{\rm YD4}$, with $\rho=-0.46$ and $p=0.03$. In contrast, the correlations of these quantities with $D_{\rm nei}$ are weak and not statistically significant. These results suggest that physical quantities related to intermediate- to long-timescale star formation histories, such as stellar mass and $\mathrm{SFR}_{100{\rm Myr}}$, are more closely related to the global protocluster structure centered on YD4 than to the distance to the nearest neighboring galaxy. This may indicate that stellar mass assembly and long-term star formation activity have proceeded more efficiently in the central region of A2744-z7p9OD.

A similar trend is seen for $A_V$. The value of $A_V$ shows significant negative correlations with $D_{\rm YD4}$ with $\rho=-0.47$ and $p=0.03$, while no significant correlation is found with $D_{\rm nei}$. As shown in the right panel of Figure~\ref{fig:Mstar_SFR_Av}, galaxies with larger stellar masses tend to have higher $A_V$ values. Therefore, the $A_V$--$D_{\rm YD4}$ trend is not necessarily independent of the $M_*$--$D_{\rm YD4}$ correlation discussed above. Nevertheless, because $A_V$ traces dust attenuation and is expected to reflect the accumulated effects of past star formation, metal enrichment, and dust build-up, the high $A_V$ values in the central region are consistent with earlier stellar mass assembly, chemical enrichment, and dust production in the protocluster core. This picture is further supported by the detection of dust continuum emission from the central galaxies YD1, YD4, and YD7W \citep{Hashimoto2023sep, Fudamoto2025oct, Umehata2025nov}, indicating that dust-rich systems are already present in the core region.

$R_{\rm e}$ also shows a significant negative correlation with $D_{\rm YD4}$, with $\rho=-0.61$ and $p=0.002$, while no significant correlation is found with $D_{\rm nei}$. Thus, galaxies closer to the protocluster center tend to be larger in rest-frame UV size. Because galaxy size generally increases with stellar mass (e.g., \citealt{Morishita2024mar}), this trend may be linked to the centrally enhanced stellar mass assembly. Nevertheless, the result suggests that the central galaxies in A2744-z7p9OD are not only more massive and dustier, but also more spatially extended than those in the outer region.

Interestingly, $\Sigma_{\rm SFR}$ shows a marginal positive correlation with $D_{\rm YD4}$ ($p\simeq0.05$), with lower $\Sigma_{\rm SFR}$ values found closer to the protocluster center. This tentative trend suggests that recent star formation in the protocluster center may not be dominated by compact starbursts, but may instead be relatively extended or less concentrated than in member galaxies in the outskirts and in typical high-redshift galaxies.

The correlations of $M_*$, $\mathrm{SFR}_{100{\rm Myr}}$, $A_V$, $R_{\rm e}$, and $\Sigma_{\rm SFR}$ with the distance from the protocluster center indicate that the central region of A2744-z7p9OD hosts more massive, dustier, and more extended galaxies with lower $\Sigma_{\rm SFR}$ than the outskirts. These trends are consistent with an inside-out growth picture, in which protocluster growth proceeds earlier in the central dominant halos (e.g., \citealt{Chiang2017aug}). In this interpretation, A2744-z7p9OD has already developed a relatively evolved core by $z\simeq7.88$. This is also consistent with \citet{Witten2025jul}, who found that the core region is more evolved than the outskirts based on SED-based star formation histories, stellar masses, and redder UV continuum slopes. 
It is also consistent with \citet{Hashimoto2023sep}, who showed, based on a comparison with the FirstLight simulations \citep{Ceverino2017sep, Nakazato2024nov}, that the galaxies in ``the Quintet'' may merge into a single massive galaxy by $z\simeq7.3$. Thus, the core of A2744-z7p9OD may already be undergoing rapid stellar mass assembly and structural growth at $z\simeq7.88$.
Accelerated galaxy formation in protocluster cores has long been discussed in both theoretical and observational studies (\citealt{Overzier2016nov,Alberts2022oct}). 
For example, massive quiescent or post-starburst galaxies have been found in the core of the SSA22 protocluster at $z=3.09$ \citep{Kubo2021sep,Umehata2025may}. A2744-z7p9OD may represent an earlier counterpart of such core-dominated systems, extending this picture to the previously unexplored regime at $z\sim8$. We note, however, that the central galaxies in A2744-z7p9OD should not be interpreted as fully quenched systems. Some galaxies may instead be in a temporary lull phase in recent star formation and could re-ignite star formation. Our results therefore suggest that, even before the formation of a quenched core, environmental differences in stellar mass assembly, dust enrichment, and star formation structure are already visible within a compact protocluster core at $z\simeq7.88$.

Overall, the member-galaxy properties examined here show stronger correlations with $D_{\rm YD4}$ than with $D_{\rm nei}$: none of the quantities is significantly correlated with $D_{\rm nei}$, whereas several show significant or marginal correlations with $D_{\rm YD4}$. This suggests that, in A2744-z7p9OD, galaxy properties are more closely linked to the global protocluster structure than to the local environment traced by the distance to the nearest neighbor. This trend contrasts with that reported for a protocluster at $z\sim3.70$, where the local environment was found to be more closely associated with galaxy properties than the global one \citep{Toshikawa2025mar}. The difference may reflect variations in the evolutionary stage and internal structure of protoclusters. A2744-z7p9OD may represent a compact, core-dominated system in which environmental effects are already imprinted on the scale of the protocluster core. Statistical studies of larger protocluster samples, enabled by future wide-field Roman surveys and deep JWST follow-up observations, will be essential to determine whether this behavior is common among protoclusters in the early Universe. We caution, however, that this result may change if future observations identify additional dusty, quiescent, or otherwise faint member galaxies currently missing from our sample.

\subsection{The Origin of Low-Ionization Gas in the Protocluster Core}
\label{subsec:The_Origin_of_Low-Ionization_Gas_in_the_Protocluster_Core}
As shown in Section~\ref{subsec:Line_Ratios}, several member galaxies of A2744-z7p9OD, particularly those in the protocluster core, exhibit low O32 values, indicating low ionization states of their ionized gas. Here, we examine the physical origin of this low-ionization gas in the dense protocluster environment.

\begin{figure*}[t]
    \centering
    \includegraphics[width=0.95\textwidth,clip]{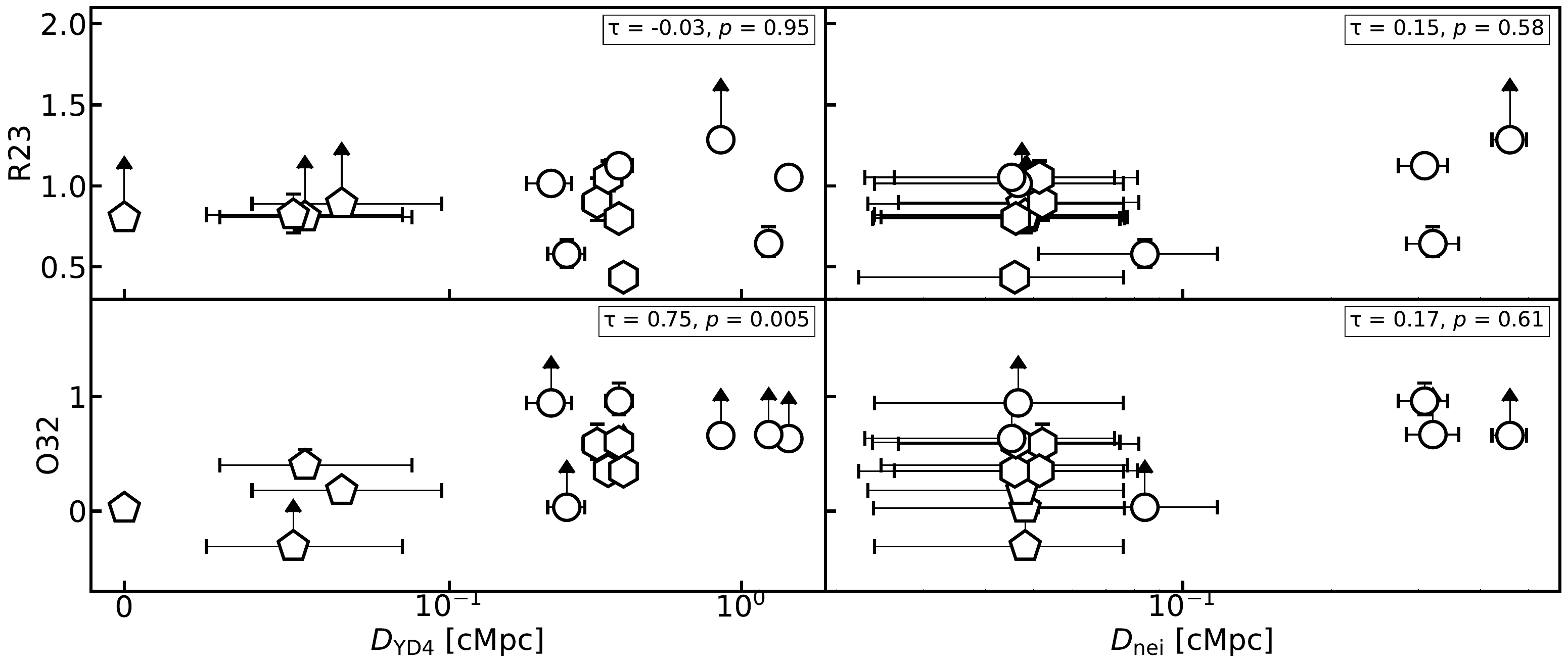}
    \caption{Line ratios (R23 and O32) of the member galaxies as a function of the two structural parameters.
    The left and right panels show $D_{\rm YD4}$ and $D_{\rm nei}$, respectively, while the top and bottom panels show R23 and O32, respectively. Based on the censored Kendall's $\tau$ test, O32 shows a significant positive correlation with $D_{\rm YD4}$, with $\tau=0.75$ and $p=0.005$, while no significant correlation is found with $D_{\rm nei}$.}
    \label{fig:distance_LineRatioproperties}
\end{figure*}

Figure~\ref{fig:distance_LineRatioproperties} shows the relations between the two structural parameters, $D_{\rm YD4}$ and $D_{\rm nei}$, and the emission-line ratios of the member galaxies. 
The bottom panels show that O32 tends to increase with increasing $D_{\rm YD4}$. To quantitatively evaluate this trend, we performed a censored Kendall's tau test \citep{Isobe1986jul}. We find that O32 shows significant positive correlations with $D_{\rm YD4}$ with $\tau=0.75$ and $p=0.005$, while no significant correlation is found between O32 and $D_{\rm nei}$. These results suggest that the low O32 values are primarily associated with the global protocluster structure rather than the local nearest-neighbor environment, indicating that gas in the central region of A2744-z7p9OD has a low ionization state.

A similar trend has been reported for another EoR protocluster. \citet{Arribas2024aug} found low O32 line ratios ($-0.2 $ to $0.6$) in the core region of SPT0311-58 at $z \sim 6.9$ compared with typical galaxies at similar redshifts. The presence of low O32 values in the cores of both A2744-z7p9OD and SPT0311-58 suggests that dense protocluster environments may host physical conditions with lower gas ionization state, although larger samples of $z>6$ protoclusters are required to determine whether this is a common property.

Because O32 primarily depends on the ionization parameter and secondarily on gas-phase metallicity, its low values can be driven by multiple galaxy properties. In the framework of \citet{Kashino2019jun}, the ionization parameter is regulated by quantities such as stellar mass, gas-phase metallicity, gas content (gas mass ratio), electron density, and sSFR (see their Figure 5). In addition, previous studies have suggested a connection between $\Sigma_{\rm SFR}$ and O32 \citep{Reddy2023aug, Reddy2023jul, Calabro2024oct}. We therefore compare O32 with these quantities, when available, to investigate the origin of the low-ionization gas in the protocluster core.

\begin{figure}[!t]
    \centering
    \includegraphics[width=\columnwidth]{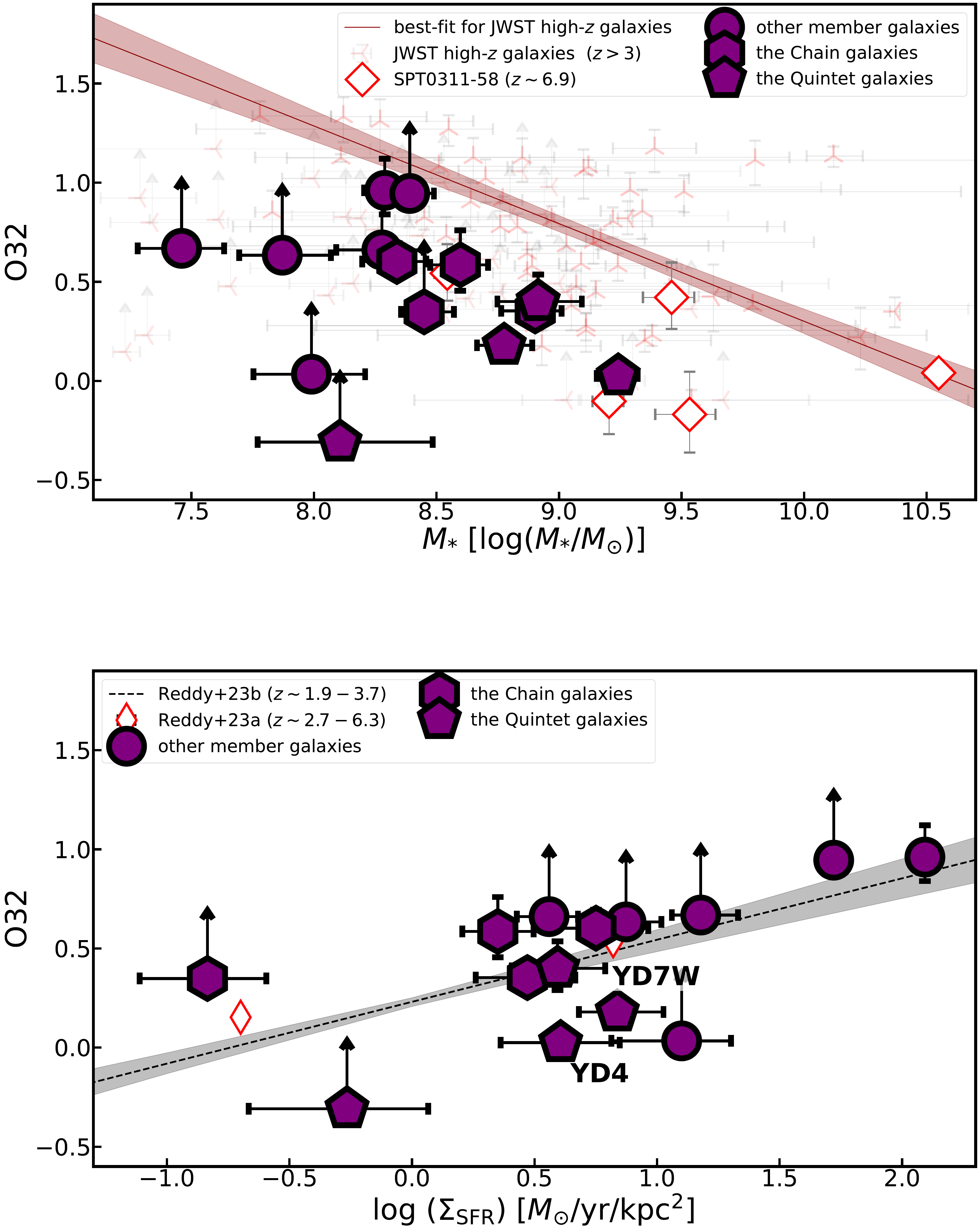}
    \caption{(Top) Stellar mass vs O32 diagram. The member galaxies of A2744-z7p9OD are shown together with individual galaxies at $z>3$ observed with JWST \citep{Cameron2023sep,Nakajima2023dec,Morishita2024aug}. The red solid line and shaded region indicate the best-fit linear relation and the $1\sigma$ scatter for the $z>3$ comparison sample, respectively. The member galaxies of A2744-z7p9OD, as well as the core galaxies of SPT0311-58 shown as red diamonds \citep{Arribas2024aug}, are located below the relation, suggesting that galaxies in protocluster cores have systematically lower O32 values than typical high-redshift galaxies with similar stellar masses. 
    (Bottom) $\log\Sigma_{\rm{SFR}}$ vs O32 diagram. The black dashed line and shaded region indicate the typical relation for star-forming galaxies at $z\sim1.9$--$3.7$ \citep{Reddy2023jul}. The red diamond represents the typical value for galaxies at $z\sim2.7$--$6.3$\citep{Reddy2023aug}.
    YD4 and YD7W lie below this relation, suggesting that their low O32 values may not be explained solely by their low $\Sigma_{\rm SFR}$.
    }
    \label{fig:O32_Mstar_SIGMASFR}
\end{figure}

\subsubsection{Stellar Mass vs O32}\label{subsubsec:M_*vsO32}
Stellar mass is a fundamental quantity that reflects the evolutionary stage of a galaxy and is closely related to the properties of ionized gas. We examine whether the low O32 values observed in the central region of A2744-z7p9OD can be explained by the stellar-mass dependence seen in typical high-redshift galaxies. 

The top panel of Figure~\ref{fig:O32_Mstar_SIGMASFR} shows the relation between stellar mass and O32 for the member galaxies of A2744-z7p9OD, together with individual galaxies at $z>3$ observed with JWST \citep{Cameron2023sep,Nakajima2023dec,Morishita2024aug}. The red solid line and shaded region indicate the best-fit relation and its $1\sigma$ scatter for the comparison sample (O32 $= (-0.5\pm0.1)\times\log(M_\ast / M_{\odot})+5.2\pm0.5$), in which O32 decreases with increasing stellar mass\footnote{This trend is consistent with those reported for other high-redshift galaxies \citep{Reddy2023jul, Umeda2026jul}.}. 

The member galaxies of A2744-z7p9OD are generally located below this best-fit relation. Three out of the five core galaxies of SPT0311-58, shown as red diamonds, are also located below the $M_\ast$--O32 relation of the comparison sample. These offsets from the typical relation suggest that the low O32 values observed in protocluster cores cannot be explained simply by their large stellar masses, and may point to significant differences in gas conditions compared with the general field galaxy population.

\subsubsection{Specific Star-Formation Rate vs O32}\label{subsubsec:sSFRvsO32}
Next, we examine the relation between sSFR and O32. In the framework of \citet{Kashino2019jun}, galaxies with higher sSFRs are expected to have a larger contribution from young massive stars and higher ionizing photon production rates on a galaxy scale. As a result, the ionizing photon flux incident on H{\sc ii} regions increases, leading to a higher ionization parameter. Therefore, if the low O32 values are caused by a decrease in sSFR, galaxies with low O32 would be expected to show lower sSFRs than typical galaxies at the same epoch. However, the sSFRs of the galaxies with low O32 in A2744-z7p9OD do not necessarily support this picture. For example, YD4 and YD7W, which show low O32 values ($0.0^{+0.1}_{-0.1}$ and $0.2^{+0.1}_{-0.1}$), have 
$\log (\mathrm{sSFR}_{10\mathrm{Myr}} / \mathrm{yr}^{-1}) = -8.0^{+0.2}_{-0.2}$ and $-7.9^{+0.2}_{-0.2}$, 
respectively. These values are slightly higher than, or at least consistent with, the typical value at the same epoch, $\log (\mathrm{sSFR}_{10\mathrm{Myr}} / \mathrm{yr}^{-1}) \sim -8.15_{-0.14}^{+0.14}$ (see also the left panel of Figure~\ref{fig:Mstar_SFR_Av}, \citealt{Simmonds2025dec}). Therefore, these galaxies do not show particularly low sSFRs. This result suggests that the low O32 values observed in the protocluster core cannot be explained solely by a decrease in the ionization parameter associated with low sSFR.

\subsubsection{Star-Formation Rate Surface Density vs O32}\label{subsubsec:SigmaSFRvsO32}
Previous studies have suggested that galaxies with higher $\Sigma_{\rm SFR}$ tend to host more compact and intense star-forming regions, which can increase the ionizing photon flux incident on the surrounding gas and lead to a higher ionization parameter \citep{Reddy2023jul, Calabro2024oct}. We therefore compare $\Sigma_{\rm SFR}$ with O32 to test whether the low O32 values in A2744-z7p9OD can be understood in terms of the $\Sigma_{\rm SFR}$--O32 relation.

The bottom panel of Figure~\ref{fig:O32_Mstar_SIGMASFR} shows the relation between $\Sigma_{\rm SFR}$ and O32 for the member galaxies of A2744-z7p9OD. For comparison, the black dashed line shows the relation for star-forming galaxies at $z\sim1.9$--$3.7$ from \citet{Reddy2023jul}\footnote{\cite{Reddy2023jul} define $\rm{O32}\equiv \log{([\mathrm{O}\text{\textsc{iii}}]4960, 5008/[\mathrm{O}\text{\textsc{ii}}]3727,3730)}$. For the comparison, we therefore applied a correction assuming [O{\sc iii}]4960 : [O{\sc iii}]5008 $=$ 1 : 2.98 \citep{Storey2000mar}}. The galaxies with moderately low O32 values, $\mathrm{O32}\sim0.5$, are broadly consistent with this relation. In contrast, galaxies with lower O32 values, such as YD4 and YD7W, lie below the relation, indicating that their O32 values are lower than expected from their measured $\Sigma_{\rm SFR}$. This suggests that less concentrated recent star formation may contribute to the low O32 values of some member galaxies, but the very low O32 values in YD4 and YD7W likely require additional factors.

We note that the comparison relation adopted here is calibrated using galaxies at $z\sim3$, whereas A2744-z7p9OD is located at $z=7.88$. Possible redshift evolution of the $\Sigma_{\rm SFR}$--O32 relation should therefore be considered (e.g., \citealt{Calabro2024oct}). It is important to establish this relation for diverse high-redshift galaxy populations and to compare protocluster members with such reference samples.

\subsubsection{Gas-Phase Metallicity vs O32}\label{subsubsec:Gas_phase_metallicityvsO32}
Gas-phase metallicity is another factor that can affect O32. In the framework of \citet{Kashino2019jun}, higher metallicity can lower the ionization parameter by absorbing the ionizing photons. High metallicity could therefore contribute to the low O32 values observed in the protocluster core.
In this study, metallicity can be probed only qualitatively using R23, because R23 is double-valued with respect to metallicity and depends on the adopted calibration\footnote{A similar caveat would apply to other strong-line metallicity indicators, such as $\hat{R}$ (e.g., \citealt{Sanders2026jun}).}. For several galaxies with low O32 values, R23 is not uniquely constrained because H$\beta$ is not detected and only lower limits on R23 are available (Figure~\ref{fig:R23_O32}). These limitations prevent us from determining whether high metallicity is the primary cause of the low O32 values. \citet{Venturi2024nov} reported that YD1 and YD4 may have slightly higher metallicities than typical galaxies at the same epoch. High metallicity therefore remains a possible contributor to the low O32 values.

\subsubsection{Electron Density vs O32}\label{subsubsec:Electron_densityvsO32}
Electron density can also regulate the ionization parameter, because it is defined as the ratio of the ionizing photon density to the hydrogen density \citep{Kashino2019jun}. In typical H{\sc ii} regions, hydrogen is almost fully ionized, and hydrogen density is approximately equal to the electron density.

We examine the electron density for YD1, YD4, YD7W, ZD6, and ZD12W, for which spectrally resolved [O{\sc ii}]3727,3730 measurements are available and at least one of the two [O{\sc ii}] lines is detected. For YD1, YD4 and YD7W, we estimate the electron density, $n_{\rm e}$, from the [O{\sc ii}]3727,3730 line ratio using the \texttt{getTemDen} function in \texttt{PyNeb} \citep{Luridiana2015jan}, assuming $T_{\rm e}=20000$ K\footnote{We also repeated the calculation assuming $T_{\rm e}=10000\ {\rm K}$, appropriate for a relatively high-metallicity environment, and confirmed that the interpretation remains unchanged.}, which is typical for galaxies at $z \sim 8$ (e.g., \citealt{Morishita2024aug}), and evaluate the uncertainties using a MC method. We obtain electron densities of $n_{\rm e}<662\ {\rm cm^{-3}}$, $n_{\rm e} =  177^{+288}_{-167}\ {\rm cm^{-3}}$ and $n_{\rm e} = 1712^{+2145}_{-950}\ {\rm cm^{-3}}$ for YD1, YD4 and YD7W, respectively. 
For ZD6 and ZD12W, we adopt the values reported by \citet{Morishita2025may}, $n_{\rm e}<40\ {\rm cm^{-3}}$ and $n_{\rm e}=2200^{+6700}_{-1600}\ {\rm cm^{-3}}$ respectively. 

Although these galaxies are all located in the central region of A2744-z7p9OD and commonly show low or moderately low O32 values ($0.0$ -- $0.6$), their electron densities span a wide range. Compared with the typical electron density expected for galaxies at $z \sim 7.88$, $n_{\rm e}\sim742\ {\rm cm^{-3}}$, derived from the relation of \citet{Abdurrouf2024sep}, YD7W is consistent within the uncertainty, while YD1, YD4 and ZD6 are lower and ZD12W may be higher, although with a large uncertainty. 
Therefore, the low O32 values in the protocluster core cannot be explained uniformly by electron density alone.

\subsubsection{Gas Reservoir vs O32}\label{subsubsec:gas_reservoirvsO32}

Finally, we discuss whether the low O32 values may be related to the neutral gas reservoir in the protocluster core. 
Here, we do not refer to the total gas content of the galaxies, but specifically to neutral gas associated with H{\sc i} gas and photodissociation regions, as traced by [C{\sc ii}] $158\mu$m emission.

In a simple H{\sc ii} region, O32 mainly reflects the ionization state of the ionized gas. However, the galaxy-integrated O32 can also be affected by the mixture of different ionized-gas components within the observing aperture. If compact H{\sc ii} regions are surrounded by optically thick neutral gas with a high H{\sc i} covering fraction, they are likely to approach ionization-bounded conditions. In this case, emission from low-ionization gas near H{\sc ii}--H{\sc i} transition layers and/or diffuse ionized gas may contribute significantly to the integrated spectrum. Since these components have lower ionization states than the highly ionized interiors of H{\sc ii} regions, they can enhance [O{\sc ii}] relative to [O{\sc iii}] and reduce the integrated O32 \citep{Pellegrini2012aug,Zhang2017apr,Sanders2017dec}.

In A2744-z7p9OD, extended [C{\sc ii}]$158\mu$m emission has been detected around ``the Quintet'', suggesting that the protocluster core contains a substantial neutral-gas reservoir \citep{Fudamoto2025oct}. Since [C{\sc ii}]$158\mu$m primarily traces neutral gas associated with photo-dissociation regions, it is not expected to be exactly co-spatial with the optical [O{\sc iii}]-emitting regions. The current data therefore do not provide direct evidence that neutral gas suppresses O32. Nevertheless, the extended [C{\sc ii}]$158\mu$m emission supports a picture in which the core of A2744-z7p9OD hosts a neutral-gas-rich, multiphase ISM. Such an ISM structure could make the galaxy-integrated line ratios closer to those expected for ionization-bounded systems, thereby contributing to the low O32 values of the member galaxies.

Taken together, the low O32 values observed in the central region of A2744-z7p9OD cannot be attributed to any single physical quantity examined in this section. This suggests that the protocluster core hosts a distinct population of low-ionization galaxies whose ISM conditions deviate from those established for field galaxies at similar epochs. Such a population may reflect physical processes specific to overdense environments that are not yet captured by existing scaling relations. Statistical observations of a larger number of protoclusters with future wide-field Roman surveys and deep JWST follow-up will be needed to determine whether this low-ionization population is common in overdense environments and to identify its physical origin.

\subsection{Implications for Cosmic Reionization}
\label{subsec:Implications_for_Cosmic_Reionization}
Based on the environmental trends and ionized-gas properties discussed above, we briefly consider the implications for cosmic reionization. Ly$\alpha$ has not been detected from the central region of A2744-z7p9OD and has only been reported from ZD4, which is located outside the core \citep{Venturi2024nov, Chen2024mar}. Together, the non-detection of Ly$\alpha$ emission from the core, the low O32 values, the possible neutral-gas reservoir traced by extended [C{\sc ii}] emission \citep{Fudamoto2025oct}, and the possible presence of high-column-density neutral hydrogen \citep{Chen2024mar} indicate that the escape of Ly$\alpha$ and ionizing photons may be suppressed in the core region of A2744-z7p9OD.
This does not necessarily contradict the general inside-out picture of reionization, in which overdense regions host abundant ionizing sources and can drive the early growth of large ionized bubbles (e.g., \citealt{Ciardi2003aug, Li2026apr}). The ionization state of the surrounding IGM reflects the cumulative ionizing output over time, and A2744-z7p9OD may have contributed to the formation of an ionized bubble in the past even if the present-day escape fraction from its core is low. Therefore, our results do not imply that the protocluster as a whole has not contributed to reionization. Rather, they suggest that the gas-rich core may not be an efficient source of escaping ionizing photons at the observed epoch, even if the surrounding large-scale environment has been ionized by the cumulative past activity of the member galaxies.

\section{Conclusions}
\label{sec:Conclusion}
We investigated the stellar populations, rest-frame UV sizes, ionized-gas properties, and internal structure of A2744-z7p9OD, a protocluster at $z=7.88$, using JWST/NIRCam imaging and JWST/NIRSpec spectroscopy. We analyzed 23 galaxies associated with A2744-z7p9OD, including 16 spectroscopically confirmed member galaxies and seven photometric member candidates. We derived stellar population properties from SED fitting, measured rest-frame UV sizes with \textsc{galfit}, and derived the emission-line ratios R23 and O32 from NIRSpec spectra. We also quantified the source-plane environments of the member galaxies using two projected structural parameters, $D_{\rm YD4}$ and $D_{\rm nei}$. Our main results are summarized as follows.

\begin{enumerate}

\item The member galaxies show diverse stellar population properties. Their stellar masses span $\log(M_\ast/M_\odot)\sim7.5$--$9.2$, and their recent star formation histories vary widely, with $\mathrm{SFR}_{10{\rm Myr}}/\mathrm{SFR}_{100{\rm Myr}}\sim0.4$--$9.2$ (Table~\ref{tab:gal_properties}). Several low-mass member galaxies show elevated recent star formation, whereas YD7E and ZD12E show relatively weak recent star formation activity. The dust attenuation spans $A_V\sim0.1$--$1.1$ mag (Figure~\ref{fig:Mstar_SFR_Av}). The total stellar mass of ``the Quintet'' is larger than that of ``the Chain'' by a factor of $\sim 4$, supporting the interpretation that ``the Quintet'' is the most massive substructure in A2744-z7p9OD.

\item The rest-frame UV effective radii of the member galaxies span $-1<\log(R_{\rm e}/{\rm kpc})<0$ and are broadly consistent with the typical size--mass relation for galaxies at similar redshifts. Their $\Sigma_{\rm SFR}$ values span $-0.8 < \log(\Sigma_{\rm SFR}{/M_{\odot}\ {\mathrm{yr}^{-1}}\ \mathrm{kpc}^{-2})} < 2.1$, and several member galaxies have lower values than the typical relation for galaxies at $6<z<10$. However, most are broadly consistent with the relation within the uncertainties, and no significant systematic offset is found for the sample as a whole (Figure~\ref{fig:Mstar_Re_SIGMASFR}).

\item The member galaxy properties are more strongly correlated with the global structural parameter $D_{\rm YD4}$ than with the local nearest-neighbor distance $D_{\rm nei}$. Stellar mass, $\mathrm{SFR}_{100{\rm Myr}}$, $A_V$, and $R_{\rm e}$ show significant negative correlations with $D_{\rm YD4}$, while $\Sigma_{\rm SFR}$ shows a marginal positive correlation (Figure~\ref{fig:distance_SEDproperties}). No significant correlations are found with $D_{\rm nei}$ for these quantities. These results suggest that stellar mass assembly, long-timescale star formation activity, dust enrichment, galaxy size growth, and the spatial concentration of recent star formation are primarily linked to the global protocluster structure in A2744-z7p9OD.

\item The member galaxies show a large galaxy-to-galaxy variation in R23, with $\Delta{\rm R23}\gtrsim0.7$ (Figure~\ref{fig:R23_O32}). If interpreted as a metallicity sequence, this variation suggests a relatively wide range of gas-phase metallicities among the member galaxies. Together with similar variations reported in the $z\sim6.9$ protocluster SPT0311-58, this result suggests that chemical enrichment may proceed in an inhomogeneous manner within dense protocluster environments already during the EoR.

\item The member galaxies also show systematically low O32 values for their epoch, indicating low ionization states. On the R23--O32 diagram (Figure~\ref{fig:R23_O32}), the lowest O32 values are found in ``the Quintet'', and O32 shows a significant positive correlation with $D_{\rm YD4}$ based on the censored Kendall's $\tau$ test ($\tau=0.75$, $p=0.005$), while no significant correlation is found with $D_{\rm nei}$ (Figure~\ref{fig:distance_LineRatioproperties}). The low O32 values cannot be explained by any single physical quantity examined in this study: stellar mass, sSFR, $\Sigma_{\rm SFR}$, metallicity, and electron density do not provide a unique explanation. The extended [C{\sc ii}] $158\,\mu$m emission around ``the Quintet'' may instead point to a complex multiphase gas structure. These results suggest that the central region of A2744-z7p9OD hosts a distinct population of low-ionization galaxies whose ISM conditions deviate from those of typical field galaxies at similar epochs.

\item The environmental trends and ionized-gas properties may have implications for cosmic reionization. Ly$\alpha$ has not been detected from the core region of A2744-z7p9OD and has only been reported from ZD4, located outside the core. Together with the low O32 values, the possible neutral-gas reservoir traced by extended [C{\sc ii}] emission, and the possible presence of high-column-density neutral hydrogen, this suggests that the current escape of Ly$\alpha$ and ionizing photons from the gas-rich core may be suppressed. However, this does not rule out a past contribution of A2744-z7p9OD to reionization, because the ionization state of the surrounding IGM reflects the cumulative ionizing output of the member galaxies.

\end{enumerate}

These results suggest that environmental effects on stellar mass assembly, dust enrichment, galaxy structure, and ionized-gas conditions were already in place in a protocluster at $z\simeq7.88$. A2744-z7p9OD appears to host an evolved, dusty, gas-rich, and low-ionization core, providing a possible example of early environmental differentiation during the EoR. Statistical observations of larger numbers of protoclusters with future wide-field Roman surveys and deep JWST follow-up will be essential to determine whether such low-ionization populations are common in overdense environments and to identify their physical origin.

\begin{acknowledgments}
We thank Nario Kuno, Shunsuke Honda, Eiichi Egami, Mitsutaka Usui, Naveen A. Reddy, Miguel Pereira-Santaella, Luca Costantin,Yuki Takagishi, Yuzuru Terui, Cassandra Barlow-Hall, Dongsheng Sun, and Shutaro Inui for helpful discussions and comments.
This work is based on observations made with the NASA/ESA/CSA James Webb Space Telescope. The data used in this work include the public data products released by the UNCOVER survey and DJA, and observations obtained through JWST programs GO-1840 and GTO-4553. The JWST data were obtained from the Mikulski Archive for Space Telescopes (MAST) at the Space Telescope Science Institute. STScI is operated by the Association of Universities for Research in Astronomy, Inc., under NASA contract NAS 5-03127.

W.O. acknowledges support from JST SPRING, Grant Number JPMJSP2124.
T.H. was supported by the Leading Initiative for Excellent Young Researchers, MEXT, Japan (HJH02007), and by JSPS KAKENHI Grant Numbers 22H01258, 23K22529, and 25K00020.
J.A.-M. acknowledges support from grants PID2024-158856NA-I00 and PID2021-127718NB-I00, funded by the Spanish Ministry of Science and Innovation/State Research Agency (MCIN/AEI/10.13039/501100011033) and by the European Regional Development Fund (ERDF),``A way of making Europe.'' J.A.-M. also acknowledges support from grant CSIC/BILATERALES2025/BIJSP25022.
L.C. and S.A. acknowledge support from grant PID2021-127718NB-I00, funded by the Spanish Ministry of Science and Innovation/State Research Agency (MCIN/AEI/10.13039/501100011033) and by the European Regional Development Fund (ERDF),``A way of making Europe.''
A.K.I. is supported by JSPS KAKENHI Grant Number 26H02069.
K.M. acknowledges support from the Waseda University Grant for Special Research Projects (Project Number 2025C-484) and JSPS KAKENHI Grant Numbers 20K14516 and 26K07139.
Y.S. is supported by JSPS KAKENHI Grant Number JP26K17200.
M.S. is supported by NASA grant 80NSSC22K1294.
D.C. is supported by research grant PID2024-156100NB-C21, financed by MICIU/AEI/10.13039/501100011033 and FEDER, EU, and by research grant CNS2024-154550, funded by MICIU/AEI/10.13039/501100011033.
M.H. is supported by JSPS KAKENHI Grant Number 22H04939.
Y.N. acknowledges support from a Flatiron Research Fellowship. The Flatiron Institute is a division of the Simons Foundation.
H.U. is supported by MEXT/JSPS KAKENHI Grant Numbers 23K20035 and 25K01039, and by the JST FOREST Program, Grant Number JP-MJFR256M.
H.Y. is supported by MEXT/JSPS KAKENHI Grant Number 26H02061 and by the JST FOREST Program, Grant Number JP-MJFR202Z.

\end{acknowledgments}

\clearpage
\onecolumngrid
\appendix
\restartappendixnumbering

\section{Lens Model}\label{subsec:Lens_Model}
We correct the lensing magnification effect as follows.
We adopt the strong-lensing (SL) model provided in UNCOVER DR4\footnote{\url{https://jwst-uncover.github.io/DR4.html\#LensingMaps}} \citep{Furtak2023aug,Price2025mar}. Below, we briefly describe the method used to construct this model (see \citealt{Furtak2023aug} for details).

This model is based on the grid-based parametric gravitational lensing method developed by \cite{Zitrin2013jan, Zitrin2013jun, Zitrin2015mar, Zitrin2021sep}, and has been further refined into an analytic form that is independent of the resolution of the input grid. The model consists of two main mass components. The first component is the cluster-member galaxies in Abell 2744, each of which is modeled as a double Pseudo-Isothermal Ellipsoid (dPIE; e.g., \citealt{Eliasdottir2007oct}). The second component is a diffuse dark-matter halo extended over the cluster as a whole, for which a Pseudo-Isothermal Elliptical Mass Distribution (PIEMD; e.g., \citealt{Jaffe1983mar}; \citealt{Keeton2001feb}) is adopted.
In this model, the SL reconstruction is built using a total of 421 cluster-member galaxies, including galaxies newly identified with JWST, and is constrained by 135 multiple images, among which 92 galaxies are spectroscopically confirmed. 

UNCOVER DR4 provides the deflection field ($\alpha$), convergence ($\kappa$), shear ($\gamma$), and lensing potential ($\psi$), all of which are normalized such that the ratio of the angular diameter distance from the lens to the source, $D_{\rm LS}$, to that from the observer to the source, $D_{\rm S}$, satisfies $D_{\rm LS}/D_{\rm S} = 1$. Using these parameters, the gravitational lensing effect of Abell 2744 can be properly corrected.

We first derive the magnification factor, $\mu$, for each member galaxy using Equation (\ref{eq:magfac}):
\begin{equation}\label{eq:magfac}
\mu = \frac{D_{\rm{LS}}}{D_{\rm{S}}} \frac{1}{|(1 - \kappa)^2 - \gamma^2|}
\end{equation}
Here, $D_{\rm{S}}$ and $D_{\rm{LS}}$ denote the angular diameter distances from the observer to A2744-z7p9OD and from the lensing cluster, Abell 2744, to A2744-z7p9OD, respectively. The resulting magnification factors are listed in Table \ref{tab:spec_obs}. We obtain $\mu \sim 1.86$--$2.76$, which is in good agreement with the values derived by \cite{Witten2025jul} assuming a source redshift of 8. 
Next, we correct for the positional offsets on the sky caused by gravitational lensing. The unlensed position, $\beta_{i}$ ($i = x, y$), that would be observed in the absence of the lens is given by the following equation:
\begin{equation}\label{equ:lensing_vec}
\beta_{i} = \theta_{i} - \frac{D_{\rm{LS}}}{D_{\rm{S}}} \alpha_{i} \ \ \ (i = x,y)
\end{equation}
where $\theta_{i}$ ($i = x, y$) is the observed position and $\alpha_{i}$ ($i = x, y$) is the deflection angle.
Using the derived source-plane coordinates, we investigate the relation between the protocluster structure and its physical properties in Sections~\ref{sec:Protocluster_Structure}, \ref{subsec:Overdensity_Environment_and_Galaxy_Properties}, and \ref{subsec:The_Origin_of_Low-Ionization_Gas_in_the_Protocluster_Core}.

\section{NIRCam photometries and NIRSpec emission-line fluxes}
Tables~\ref{tab:Observed_measurements_table1}, \ref{tab:Observed_measurements_table2}, and \ref{tab:Observed_measurements_table3} present the NIRCam photometries and NIRSpec emission-line fluxes used in this study. The NIRCam photometries were measured in the 19 bands described in Section~\ref{subsec:NIRCam_Imaging_Data}, and the NIRSpec line fluxes were measured from the IFS and MOS spectra as described in Section~\ref{subsubsec:Line_Flux_Measurement}. 
For [O{\sc ii}], we report the total doublet flux, [O{\sc ii}]$3727,3730$. For high-resolution spectra in which the doublet is resolved, we also list the individual [O{\sc ii}]$3727$ and [O{\sc ii}]$3730$ fluxes measured from double-Gaussian fits. For PRISM spectra, only the total [O{\sc ii}]$3727,3730$ flux is reported. We regard [O{\sc ii}] as detected when the summed doublet flux has $\mathrm{SNR}>3$. All values in the table are observed quantities and are not corrected for gravitational lensing magnification.

\setlength{\tabcolsep}{3pt}
\begin{deluxetable}{lccccccccc}
\tabletypesize{\scriptsize}
\tablecaption{Observed NIRCam photometries and NIRSpec emission-line fluxes for YD1, YD4, YD6, YD7W, YD7E, s1, ZD1, ZD3, and ZD6. The NIRCam photometries are given in units of $100\ {\rm nJy}$, and the NIRSpec emission-line fluxes are given in units of $10^{-18}\ {\rm erg\ s^{-1}\ cm^{-2}}$. Upper limits on the emission-line fluxes are $3\sigma$. These values are not corrected for gravitational lensing magnification.}
\label{tab:Observed_measurements_table1}
\tablehead{
\colhead{}& \colhead{YD1} & \colhead{YD4} & \colhead{YD6} & \colhead{YD7W} & \colhead{YD7E} & \colhead{s1} & \colhead{ZD1} & \colhead{ZD3} & \colhead{ZD6}
}
\startdata
$f_{\nu,\rm{F070W}}$ & $0.00\pm0.13$&$0.01\pm0.12$&$-0.03\pm0.11$&$0.01\pm0.06$&$-0.05\pm0.10$&$-0.01\pm0.04$&$-0.02\pm0.04$&$-0.02\pm0.10$&$-0.10\pm0.13$    \\
$f_{\nu,\rm{F090W}}$& $-0.02\pm0.12$&$0.09\pm0.12$&$-0.04\pm0.10$&$-0.01\pm0.05$&$0.03\pm0.09$&$0.02\pm0.03$&$-0.03\pm0.04$&$-0.03\pm0.08$&$-0.03\pm0.12$ \\
$f_{\nu,\rm{F115W}}$& $0.16\pm0.16$&$0.19\pm0.15$&$0.22\pm0.13$&$0.18\pm0.07$&$0.42\pm0.12$&$0.04\pm0.05$&$0.05\pm0.05$&$0.47\pm0.11$&$0.36\pm0.15$   \\
$f_{\nu,\rm{F140M}}$&  $0.35\pm0.32$&$0.68\pm0.32$&$0.43\pm0.28$&$0.16\pm0.14$&$0.80\pm0.23$&$0.12\pm0.09$&$0.08\pm0.10$&$0.48\pm0.23$&$0.37\pm0.31$   \\
$f_{\nu,\rm{F150W}}$&  $0.34\pm0.16$&$0.56\pm0.16$&$0.55\pm0.14$&$0.40\pm0.08$&$0.79\pm0.13$&$0.07\pm0.05$&$0.11\pm0.06$&$0.66\pm0.12$&$0.66\pm0.16$   \\
$f_{\nu,\rm{F182M}}$& $0.15\pm0.25$&$0.91\pm0.24$&$0.55\pm0.18$&$0.38\pm0.12$&$0.78\pm0.18$&$0.09\pm0.07$&$0.09\pm0.08$&$0.46\pm0.17$&$0.47\pm0.23$   \\
$f_{\nu,\rm{F200W}}$&  $0.42\pm0.22$&$0.81\pm0.18$&$0.53\pm0.20$&$0.48\pm0.09$&$0.83\pm0.16$&$0.11\pm0.06$&$0.12\pm0.08$&$0.59\pm0.13$&$0.65\pm0.18$ \\
$f_{\nu,\rm{F210M}}$& $0.18\pm0.25$&$0.84\pm0.25$&$0.46\pm0.20$&$0.33\pm0.11$&$0.87\pm0.17$&$0.00\pm0.07$&$0.05\pm0.08$&$0.49\pm0.18$&$0.46\pm0.22$    \\
$f_{\nu,\rm{F250M}}$& $0.46\pm0.42$&$0.95\pm0.54$&$0.60\pm0.30$&$0.35\pm0.26$&$0.73\pm0.31$&$0.02\pm0.14$&$0.17\pm0.17$&$0.42\pm0.30$&$0.32\pm0.35$   \\
$f_{\nu,\rm{F277W}}$&  $0.49\pm0.21$&$1.00\pm0.15$&$0.66\pm0.15$&$0.51\pm0.11$&$0.84\pm0.15$&$0.09\pm0.05$&$0.09\pm0.07$&$0.52\pm0.13$&$0.68\pm0.12$   \\
$f_{\nu,\rm{F300M}}$&  $0.56\pm0.15$&$1.14\pm0.15$&$0.60\pm0.13$&$0.54\pm0.08$&$0.73\pm0.13$&$0.05\pm0.05$&$0.08\pm0.06$&$0.43\pm0.12$&$0.45\pm0.13$   \\
$f_{\nu,\rm{F335M}}$& $0.85\pm0.12$&$1.57\pm0.12$&$0.86\pm0.10$&$0.72\pm0.06$&$1.18\pm0.09$&$0.07\pm0.04$&$0.14\pm0.05$&$0.67\pm0.09$&$0.84\pm0.12$    \\
$f_{\nu,\rm{F356W}}$& $0.88\pm0.07$&$1.53\pm0.07$&$0.93\pm0.06$&$0.75\pm0.04$&$1.25\pm0.06$&$0.11\pm0.03$&$0.14\pm0.03$&$0.70\pm0.06$&$1.01\pm0.08$    \\
$f_{\nu,\rm{F360M}}$&  $0.63\pm0.12$&$1.57\pm0.12$&$0.87\pm0.10$&$0.66\pm0.06$&$1.29\pm0.09$&$0.08\pm0.04$&$0.24\pm0.05$&$0.63\pm0.09$&$0.84\pm0.11$   \\
$f_{\nu,\rm{F410M}}$&  $0.67\pm0.09$&$1.48\pm0.09$&$0.83\pm0.08$&$0.72\pm0.05$&$1.32\pm0.08$&$0.08\pm0.03$&$0.15\pm0.04$&$0.63\pm0.07$&$0.87\pm0.10$   \\
$f_{\nu,\rm{F430M}}$& $1.06\pm0.18$&$1.71\pm0.17$&$0.98\pm0.15$&$0.91\pm0.09$&$1.33\pm0.14$&$0.06\pm0.06$&$0.17\pm0.07$&$0.67\pm0.13$&$1.17\pm0.17$    \\
$f_{\nu,\rm{F444W}}$&  $1.37\pm0.09$&$1.93\pm0.09$&$1.09\pm0.08$&$1.02\pm0.05$&$1.42\pm0.07$&$0.23\pm0.03$&$0.16\pm0.04$&$0.87\pm0.07$&$1.28\pm0.09$   \\
$f_{\nu,\rm{F460M}}$&  $0.91\pm0.22$&$1.33\pm0.21$&$0.86\pm0.19$&$0.47\pm0.12$&$1.19\pm0.17$&$0.03\pm0.08$&$0.09\pm0.09$&$0.56\pm0.17$&$0.68\pm0.22$   \\
$f_{\nu,\rm{F480M}}$&  $0.60\pm0.27$&$1.17\pm0.27$&$0.81\pm0.23$&$0.50\pm0.15$&$1.23\pm0.22$&$-0.10\pm0.09$&$0.03\pm0.11$&$0.56\pm0.20$&$0.90\pm0.27$  \\
$F_{\rm{[OII]3727}}$  &$< 0.78$&$1.35\pm0.34$& - &$1.13\pm0.20$& - &$< 0.53$& - &$< 0.32$&$0.61\pm0.10$  \\
$F_{\rm{[OII]3730}}$
&$1.08\pm0.30$&$2.31\pm0.35$& - &$0.83\pm0.18$& - &$< 0.53$& - &$< 0.32$&$0.79\pm0.10$  \\
$F_{\rm{[OII]3727,3730}}$ &$1.55\pm0.39$&$3.64\pm0.48$& - &$1.97\pm0.29$& - &$< 1.02$& - &$0.64\pm0.21$&$1.40\pm0.14$  \\
$F_{\rm{H}\beta}$ &$< 0.81$&$< 1.07$& - &$< 0.64$& - &$0.71\pm0.19$& - &$0.51\pm0.14$&$0.58\pm0.12$  \\
$F_{\rm{[OIII]5008}}$&$6.03\pm0.27$&$5.40\pm0.33$& - &$4.01\pm0.19$& - &$2.85\pm0.20$& - &$2.58\pm0.10$&$3.74\pm0.11$  \\
\enddata
\end{deluxetable}

\setlength{\tabcolsep}{4pt}
\begin{deluxetable}{lccccccccc}
\tabletypesize{\scriptsize}
\tablecaption{Same as Table~\ref{tab:Observed_measurements_table1}, but for the member galaxies ZD12E, ZD12W, ZD2, ZD4, GLASSz8-2, YD8, PC2E, PC5, and ALT-41799.}
\label{tab:Observed_measurements_table2}
\tablehead{
\colhead{}& \colhead{ZD12E} & \colhead{ZD12W} & \colhead{ZD2} & \colhead{ZD4} & \colhead{GLASSz8-2} & \colhead{YD8} & \colhead{PC2E} & \colhead{PC5} & \colhead{ALT-41799}
}
\startdata
$f_{\nu,\rm{F070W}}$ & $0.02\pm0.02$&$-0.01\pm0.05$&$-0.06\pm0.14$&$0.01\pm0.03$&$0.03\pm0.11$&$0.01\pm0.13$&$-0.01\pm0.05$&$0.03\pm0.03$&$0.12\pm0.08$     \\
$f_{\nu,\rm{F090W}}$& $0.01\pm0.01$&$-0.03\pm0.04$&$-0.02\pm0.13$&$-0.00\pm0.02$&$-0.00\pm0.11$&$0.03\pm0.10$&$0.01\pm0.04$&$0.04\pm0.03$&$0.28\pm0.06$ \\
$f_{\nu,\rm{F115W}}$& $0.22\pm0.02$&$0.14\pm0.06$&$0.92\pm0.18$&$0.06\pm0.04$&$0.73\pm0.15$&$0.52\pm0.16$&$0.18\pm0.07$&$0.15\pm0.05$&$1.08\pm0.10$   \\
$f_{\nu,\rm{F140M}}$&  $0.30\pm0.04$&$0.19\pm0.11$&$1.62\pm0.35$&$0.09\pm0.07$&$0.97\pm0.29$&$0.80\pm0.25$&$0.39\pm0.13$&$0.12\pm0.08$&$1.37\pm0.17$   \\
$f_{\nu,\rm{F150W}}$&  $0.29\pm0.02$&$0.28\pm0.06$&$1.34\pm0.19$&$0.12\pm0.04$&$1.11\pm0.18$&$0.77\pm0.17$&$0.33\pm0.08$&$0.23\pm0.05$&$1.72\pm0.11$   \\
$f_{\nu,\rm{F182M}}$& $0.30\pm0.03$&$0.20\pm0.08$&$1.35\pm0.26$&$0.04\pm0.05$&$0.95\pm0.22$&$0.57\pm0.21$&$0.28\pm0.09$&$0.21\pm0.06$&$1.46\pm0.14$    \\
$f_{\nu,\rm{F200W}}$&  $0.26\pm0.03$&$0.27\pm0.07$&$1.36\pm0.22$&$0.12\pm0.04$&$0.99\pm0.23$&$0.67\pm0.23$&$0.32\pm0.09$&$0.24\pm0.06$&$1.93\pm0.17$ \\
$f_{\nu,\rm{F210M}}$& $0.23\pm0.03$&$0.25\pm0.08$&$1.39\pm0.26$&$0.11\pm0.05$&$1.00\pm0.22$&$0.59\pm0.23$&$0.30\pm0.09$&$0.26\pm0.05$&$1.84\pm0.15$    \\
$f_{\nu,\rm{F250M}}$& $0.25\pm0.04$&$0.19\pm0.12$&$1.36\pm0.34$&$0.10\pm0.12$&$0.96\pm0.45$&$0.61\pm0.52$&$0.27\pm0.27$&$0.17\pm0.15$&$1.61\pm0.35$    \\
$f_{\nu,\rm{F277W}}$&  $0.24\pm0.02$&$0.25\pm0.05$&$1.41\pm0.13$&$0.10\pm0.03$&$1.07\pm0.15$&$0.68\pm0.18$&$0.33\pm0.09$&$0.22\pm0.05$&$1.59\pm0.12$    \\
$f_{\nu,\rm{F300M}}$&  $0.24\pm0.02$&$0.20\pm0.06$&$1.56\pm0.15$&$0.11\pm0.03$&$1.07\pm0.15$&$0.71\pm0.20$&$0.36\pm0.09$&$0.24\pm0.06$&$1.45\pm0.13$   \\
$f_{\nu,\rm{F335M}}$& $0.31\pm0.02$&$0.25\pm0.05$&$1.80\pm0.13$&$0.13\pm0.03$&$1.40\pm0.12$&$0.87\pm0.14$&$0.41\pm0.06$&$0.21\pm0.04$&$1.70\pm0.10$    \\
$f_{\nu,\rm{F356W}}$& $0.34\pm0.01$&$0.35\pm0.03$&$1.62\pm0.08$&$0.13\pm0.02$&$1.25\pm0.07$&$0.82\pm0.08$&$0.40\pm0.04$&$0.19\pm0.03$&$1.75\pm0.05$    \\
$f_{\nu,\rm{F360M}}$&  $0.32\pm0.02$&$0.27\pm0.05$&$1.64\pm0.13$&$0.13\pm0.03$&$1.20\pm0.11$&$0.83\pm0.13$&$0.41\pm0.05$&$0.16\pm0.04$&$1.69\pm0.08$   \\
$f_{\nu,\rm{F410M}}$&  $0.37\pm0.02$&$0.27\pm0.04$&$1.41\pm0.11$&$0.14\pm0.03$&$1.01\pm0.10$&$0.63\pm0.12$&$0.37\pm0.05$&$0.13\pm0.03$&$1.72\pm0.07$   \\
$f_{\nu,\rm{F430M}}$& $0.35\pm0.03$&$0.45\pm0.08$&$2.76\pm0.19$&$0.27\pm0.05$&$2.04\pm0.18$&$0.97\pm0.18$&$0.62\pm0.10$&$0.21\pm0.06$&$2.66\pm0.14$    \\
$f_{\nu,\rm{F444W}}$&  $0.39\pm0.01$&$0.51\pm0.04$&$3.55\pm0.10$&$0.18\pm0.02$&$2.73\pm0.10$&$1.41\pm0.11$&$0.80\pm0.04$&$0.31\pm0.04$&$3.33\pm0.08$   \\
$f_{\nu,\rm{F460M}}$&  $0.38\pm0.04$&$0.36\pm0.10$&$0.88\pm0.24$&$0.16\pm0.06$&$0.72\pm0.20$&$0.60\pm0.23$&$0.08\pm0.11$&$0.09\pm0.08$&$1.82\pm0.16$    \\
$f_{\nu,\rm{F480M}}$&  $0.29\pm0.05$&$0.33\pm0.11$&$0.65\pm0.28$&$0.04\pm0.07$&$0.61\pm0.26$&$0.60\pm0.27$&$0.26\pm0.13$&$0.11\pm0.09$&$1.61\pm0.18$   \\
$F_{\rm{[OII]3727}}$  &$< 0.32 $&$0.41\pm0.11$&$< 0.67$& - & - & - & - & - & -   \\
$F_{\rm{[OII]3730}}$
&$< 0.32$& $0.38\pm0.11$& $< 0.67$& - & - & - & - & - & -   \\
$F_{\rm{[OII]3727,3730}}$ &$< 0.44$&$0.79\pm0.17$&$<1.79$&$< 0.36$&$1.37\pm0.4$&$< 1.36$&$< 0.95$&$< 0.46$& -   \\
$F_{\rm{H}\beta}$ &$1.01\pm0.14$&$0.92\pm0.13$&$2.31\pm0.15$&$0.35\pm0.05$&$1.59\pm0.14$&$< 0.88$&$0.89\pm0.14$&$0.96\pm0.22$ & -   \\
$F_{\rm{[OIII]5008}}$&
$1.76\pm0.11$&$3.74\pm0.11$&$17.5\pm0.2$&$0.92\pm0.06$&$15.10\pm0.13$&$9.95\pm0.21$&$7.14\pm0.12$&$3.10\pm0.07$& -  \\
\enddata
\end{deluxetable}

\setlength{\tabcolsep}{4pt}
\begin{deluxetable}{lccccccccc}
\tabletypesize{\scriptsize}
\tablecaption{Same as Table~\ref{tab:Observed_measurements_table1}, but for the member galaxies PC1, PC2W, PC3, PC4, and PC6.}
\label{tab:Observed_measurements_table3}
\tablehead{
\colhead{}& \colhead{PC1} & \colhead{PC2W} & \colhead{PC3} & \colhead{PC4} & \colhead{PC6}
}
\startdata
$f_{\nu,\rm{F070W}}$ & $0.01 \pm 0.06$&$0.02 \pm 0.07$&-&$0.01 \pm 0.05$&$0.05 \pm 0.04$     \\
$f_{\nu,\rm{F090W}}$ & $0.02 \pm 0.06$&$0.06 \pm 0.05$&$0.08\pm0.07$&$0.01 \pm 0.04$&$0.03 \pm 0.04$     \\
$f_{\nu,\rm{F115W}}$ & $0.10 \pm 0.07$&$0.42 \pm 0.09$&$0.48\pm0.07$&$0.21 \pm 0.05$&$0.05 \pm 0.04$     \\
$f_{\nu,\rm{F140M}}$ & $0.17 \pm 0.16$&$0.75 \pm 0.24$&-&$0.16 \pm 0.12$&$0.19 \pm 0.13$     \\
$f_{\nu,\rm{F150W}}$ & $0.14 \pm 0.07$&$0.65 \pm 0.09$&$0.47\pm0.1$&$0.21 \pm 0.05$&$0.07 \pm 0.05$     \\
$f_{\nu,\rm{F182M}}$ & $0.18 \pm 0.13$&$0.63 \pm 0.20$&-&$0.2 \pm 0.09$&$0.03 \pm 0.11$     \\
$f_{\nu,\rm{F200W}}$ & $0.18 \pm 0.08$&$0.59 \pm 0.10$&$0.33\pm0.08$&$0.18 \pm 0.05$&$0.08 \pm 0.06$     \\
$f_{\nu,\rm{F210M}}$ & $0.12 \pm 0.14$&$0.54 \pm 0.23$&-&$0.14 \pm 0.11$&$0.06 \pm 0.13$     \\
$f_{\nu,\rm{F250M}}$ & $0.16 \pm 0.11$&$0.56 \pm 0.16$&-&$0.17 \pm 0.08$&$0.10 \pm 0.09$     \\
$f_{\nu,\rm{F277W}}$ & $0.16 \pm 0.06$&$0.55 \pm 0.09$&$0.47\pm0.05$&$0.15 \pm 0.04$&$0.08 \pm 0.06$     \\
$f_{\nu,\rm{F300M}}$ & $0.11 \pm 0.09$&$0.54 \pm 0.14$&-&$0.19 \pm 0.07$&$0.05 \pm 0.07$     \\
$f_{\nu,\rm{F335M}}$ & $0.14 \pm 0.08$&$0.65 \pm 0.10$&-&$0.19 \pm 0.06$&$0.13 \pm 0.06$     \\
$f_{\nu,\rm{F356W}}$ & $0.18 \pm 0.04$&$0.62 \pm 0.07$&$0.49\pm0.05$&$0.16 \pm 0.04$&$0.08 \pm 0.05$     \\
$f_{\nu,\rm{F360M}}$ & $0.19 \pm 0.06$&$0.64 \pm 0.07$&-&$0.18 \pm 0.05$&$0.08 \pm 0.04$     \\
$f_{\nu,\rm{F410M}}$ & $0.14 \pm 0.05$&$0.64 \pm 0.07$&$0.40\pm0.06$&$0.14 \pm 0.04$&$0.07 \pm 0.05$     \\
$f_{\nu,\rm{F430M}}$ & $0.34 \pm 0.11$&$0.73 \pm 0.16$&-&$0.28 \pm 0.09$&$0.12 \pm 0.09$     \\
$f_{\nu,\rm{F444W}}$ & $0.25 \pm 0.05$&$0.87 \pm 0.06$&$0.91\pm0.05$&$0.27 \pm 0.04$&$0.14 \pm 0.04$     \\
$f_{\nu,\rm{F460M}}$ & $0.08 \pm 0.13$&$0.56 \pm 0.18$&-&$0.11 \pm 0.11$&$0.07 \pm 0.10$     \\
$f_{\nu,\rm{F480M}}$ & $0.00\pm0.00$&$0.70 \pm 0.15$&$0.25\pm0.17$&$0.09 \pm 0.11$&$0.07 \pm 0.09$     \\
$F_{\rm{[OII]3727}}$ & - & - & - & - & -  \\
$F_{\rm{[OII]3730}}$ & - & - & - & - & -  \\
$F_{\rm{[OII]3727,3730}}$ & - & - & - & - & -  \\
$F_{\rm{H}\beta}$     & - & - & - & - & -  \\
$F_{\rm{[OIII]5008}}$ & - & - & - & - & -  \\
\enddata
\end{deluxetable}

\onecolumngrid
\section{SED fitting for member galaxies}
\begin{figure*}[!h]
    \centering
    \includegraphics[width=1\textwidth,clip]{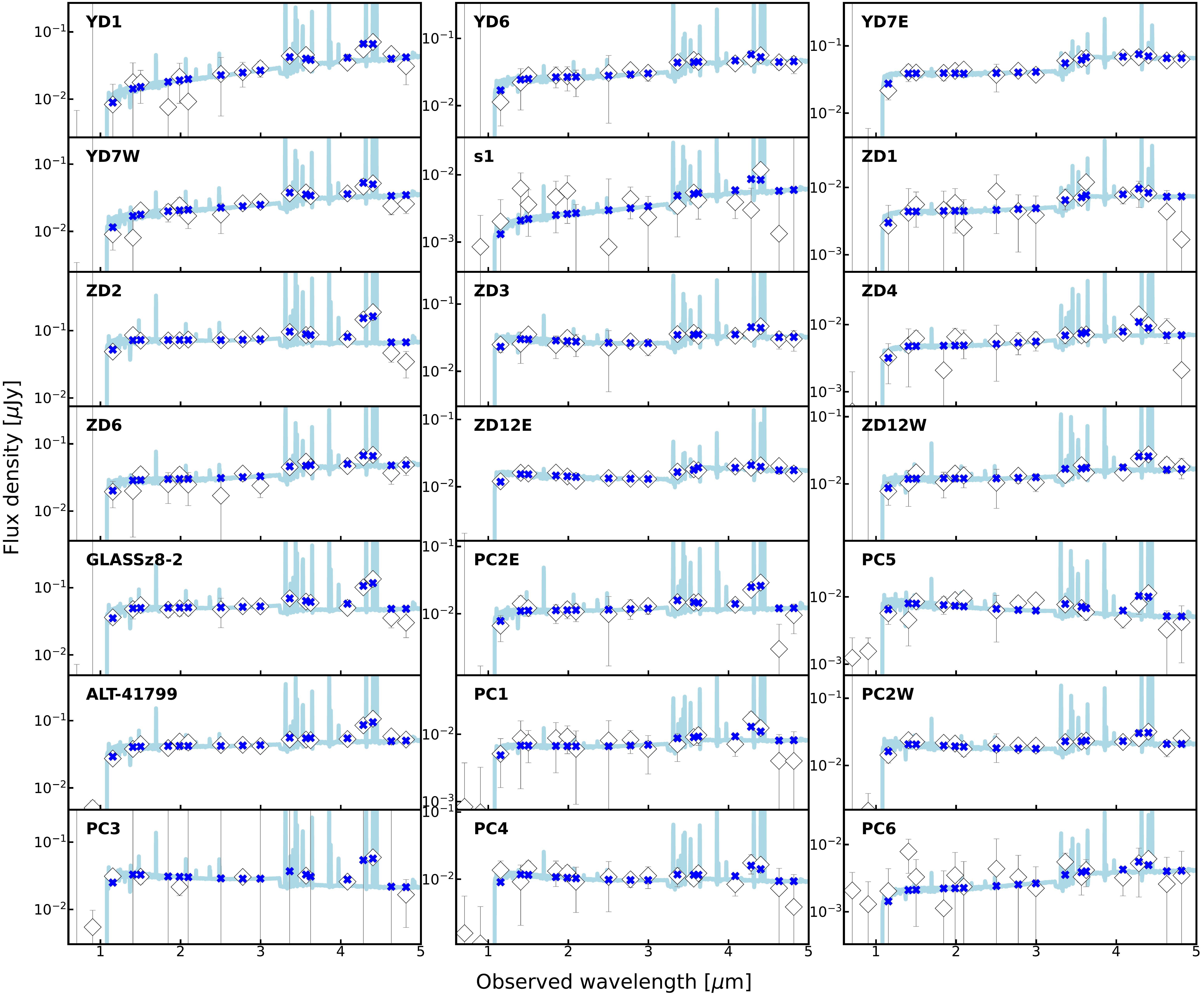}
    \caption{Same as Figure~\ref{fig:YD4_YD8_model_SED}, but for the other member galaxies. A non-parametric SFH model was adopted for all galaxies except PC3, for which a delayed-$\tau$ model was used.}
    \label{fig:memgal_model_SED}
\end{figure*}

Figure~\ref{fig:memgal_model_SED} shows the model SEDs for the member galaxies of A2744-z7p9OD. The observed NIRCam photometries are compared with the best-fit \texttt{BAGPIPES} model SEDs obtained using the procedure described in Section~\ref{sec:SED_Fitting}. We adopted a non-parametric SFH model for all galaxies except PC3, for which a delayed-$\tau$ model was used. This figure demonstrates that the best-fit models generally reproduce the observed photometries within the uncertainties, while the constraints are weaker for faint sources or sources with non-detections in several bands.

\clearpage
\twocolumngrid

\clearpage

\facilities{JWST(NIRCam, NIRSpec)}

\software{astropy \citep{Astropy_Collaboration2022aug},
          photutils \citep{Bradley2023may}, 
          scipy  \citep{Virtanen2020feb}, 
          pypher  \citep{Boucaud2016dec}, 
          BAGPIPES \citep{Carnall2018nov},
          GALFIT \citep{Peng2002jul, Peng2010jun},
          PyNeb \citep{Luridiana2015jan},
          }

% \bibliography{ref}{}
\bibliographystyle{aasjournalv7}

\end{document}